\documentclass[12pt]{article}
\usepackage{listings,upquote,verbatim,amsmath,cite,epsfig,bm,wrapfig,wasysym,amssymb,axodraw,graphics,graphicx,color,appendix,multirow}
\usepackage[abs]{overpic}
\usepackage[hypertex]{hyperref}

\hypersetup
{
	colorlinks,bookmarksopen,bookmarksnumbered,
	linkcolor=blue,citecolor=blue,urlcolor=blue,
	linkbordercolor={0 0 1}, urlbordercolor={0 0 1},
	pdfstartview=FitH, plainpages=false,
	pdfauthor = {Noam Hod},
	pdftitle = {An introduction to the Moses project},
	pdfsubject = {Search for Kaluza-Klein excitations at the LHC},
	pdfkeywords = {Kaluza-Klein, Z', MCnet, Moses, ATLAS, LHC, Monte Carlo, C++, Pythia8},
	pdfcreator = {LaTeX with hyperref package},
	pdfproducer = {dvips + ps2pdf}
}

\newcommand{\shellfont}[1]{{\fontfamily{pcr}\selectfont\small#1}}

\newcommand{\mosesWithSpace}[0]{{\sc{Moses }}}
\newcommand{\mosesNoSpace}[0]{{\sc{Moses}}}

\begin{document}

\begin{titlepage}

\begin{center}

{\LARGE\textbf{Introduction to the MCnet}} {\LARGE\textbf{M}}{\large\textbf{OSES}} {\LARGE\textbf{project}}\\
{\LARGE\textbf{and Heavy gauge bosons search at the LHC}}
\vspace{0.5cm}

{\large G}{\small IDEON} {\large B}{\small ELLA}, {\large E}{\small REZ} {\large E}{\small TZION},
{\large N}{\small OAM} {\large H}{\small OD}\footnote{email: noam.hod@cern.ch}\\
{\it The Raymond and Beverly Sackler School of Physics \& Astronomy,\\ Tel-Aviv University}\\
\vspace{0.7cm}
{\large M}{\small ARK} {\large S}{\small UTTON}\footnote{email: sutt@cern.ch}\\
{\it The Department of Physics and Astronomy,\\ University of Sheffield}\\
\vspace{0.4cm}

\end{center}

\begin{abstract}
\noindent
This is a technical document that provides supporting information and details of the publicly available code used for the preparation of the analysis for preprint ``{\it A search for heavy Kaluza-Klein electroweak gauge bosons at the LHC}'' (submitted to JHEP).\\\\
\noindent
The \mosesWithSpace C++ framework is a project written for probing and developing new models for High Energy Physics processes which allows complete events to be simulated by interface with the standard simulation program Pythia8~\cite{PYTHIA8}. 
This paper demonstrates the usage of \mosesWithSpace in a study of the nature of Kaluza-Klein (KK) excitations in a specific model where the $SU(2) \times U(1)$ gauge fields can exist in a single Extra Dimension (ED) compactified on a $S^1/Z_2$ orbifold, while the matter fermions and $SU(3)$ gauge fields are localized in the $3d$-brane.
Using this framework, the events have been fully simulated at hadron level including initial and final state radiation.
The study of particle decays was used to develop a method to distinguish between this Kaluza-Klein model and processes with similar final states.
As a consequence, the possibility of observing and identifying a signal of the first excited KK state of the $\gamma/Z^0$ bosons in the LHC is also discussed.
\end{abstract}
\vspace{0.2cm}

\centering{\today}

\end{titlepage}

\tableofcontents

\section{About the project}
The \mosesWithSpace framework was developed during an MCnet 4-months project at University College London.
Subsequently further development took place in Tel Aviv university. 
This paper discusses the first proof-of-concept version. 
The code has been developed in C++ using standard Gnu development tools running on Scientific Linux. 
It requires the packages ROOT\cite{ROOT}, LHAPDF\cite{LHAPDF}, HepPDT\cite{HEPPDT} and Pythia8 to be accessible.
This paper contains two parts: {\em (a)} the development of the validation processes, and {\em (b)} the specific implementation of the KK process and the corresponding analysis of the events at the generator level.

\clearpage

\section{Introduction}
\subsection{Heavy Kaluza-Klein gauge bosons search at the LHC}
The LHC is expected to be ready for colliding beams at the end of 2009.
It is designed to collide proton beams at 14 TeV, the highest CM energy ever reached in a laboratory.
It will greatly enlarge the kinematic region for the search for physics phenomena Beyond the Standard Model (BSM).
Several BSM theories predict the existence of other dimensions in addition to the usual three spatial and one time dimension.
These models allow various particles to propagate into the extra-dimensional bulk.
The TeV$^{-1}$ ED model considered here allows the KK modes of $SU(2)\times U(1)$ gauge fields to propagate into the extra-dimensional bulk while restricting all the matter fermions and the $SU(3)$ gauge fields to be localized in the usual 3$d$ brane~\cite{ANTONIADIS,ARKANIHAMED,ARKANIHAMED2}.
One objective of the LHC program in the context of these models is to search for a signal of the first excited KK mode of the $SU(2) \times U(1)$ gauge fields, denoted by $\gamma^*$ and $Z^*$.

A popular model that does not involve extra dimensions is also considered for comparison.
An extra heavy boson arising from the breaking of the $E_6$ group~\cite{ZPRIMEPHENOMENOLOGY,ZPRIMEPHYSICS} is assumed.
The signal of this extra heavy boson, denoted by $Z'$, can demonstrate similar characteristics to the KK signal for which a technique to distinguish between the two models is required.

Another popular possibility that is \textbf{not} considered in this paper is the RS~\cite{RANDALSANDRUM} model that predicts the existence of Kaluza-Klein spin-2 gravitons, denoted by $G$.
In some models, the mass of first KK graviton excitation can be the same as the mass of the spin-1 $\gamma^*/Z^*$ or $Z'$, and therefore, there is a need to identify the $G$ signature and especially its spin.

In this paper, the signature of the first two cases is studied where the produced bosonic candidates from either the Standard Model (SM) $\gamma/Z^0$ bosons, the heavy KK $\gamma^*/Z^*$ bosons, or the extra bosons $Z'$, decay all into charged leptons.

This KK model is particularly interesting because of the strong destructive interference that manifests itself at much lower invariant masses comparing to the resonance itself.
In this model this will always occur for masses around half of the resonance mass.
For example, with a resonance at 4~TeV, this will occur around 2~TeV.
Therefore, at the LHC, the suppression of the cross section can be observed much earlier than the resonance itself.
This will not happen for the various $Z'$ possible signals that go along with the SM Z-line shape up to masses near the $Z'$ resonance~\cite{RIZZOINTERFERENCE}.

Powerful methods to quantify the sensitivity for BSM physics at the $\sim$1 TeV scale are presented.
For an observed resonance above a small SM background around $\sim$4 TeV, a measurement of the charged lepton kinematic distributions can provide the discrimination between the two spin-1 BSM models presented in this paper.
The sensitivity for this discrimination depends on the LHC luminosity and the collisions energy, and it should be possible already with an integrated luminosity of $\mathcal{L}$=100 fb$^{-1}$ by applying the Kolmogorov test on the invariant mass and the angular distributions\footnote{The Kolmogorov test is applied between the pseudo-data samples and the MC reference samples.}.
This measurement can also quantify some interesting physical aspects of the new models and support the discrimination between them to some extent.
The main observable that is extracted, in this context, is the forward backward asymmetry\footnote{There is a big difference between the KK and $Z'$ asymmetry where the KK behave similar to the SM.} of the angular distributions.
Although it is \textbf{not} shown here, this measurement can enable the classification of the resonance as spin-1 or, for instance, spin-2 to some extent\footnote{For that purpose, the full angular distribution, of both decay angles, can be used\cite{AZIMUTHALRS}.}.
Further consistent experimental studies on $Z'$, $\gamma^*/Z^*$ and the $G$ possible signals in the LHC can be found elsewhere~\cite{ZPRIMETOE+E-,KKEXPERIMENTAL,GRAVITON,GRAVITON3}.

\subsection{Project overview}
Many processes are implemented in standard event generators, however, there are numerous specific BSM processes which are not modeled.
This framework can be used to integrate various new BSM processes with the standard event generator Pythia8.
New processes can of course be directly implemented in Pythia8 where there is a special interface class to that job.
In some cases it is natural and convenient to integrate the new model.
In other cases, it may turn out to be more complex.
In that light, the \mosesWithSpace framework has some advantages.
It allows, for instance, to utilize the interfaces with the standard HepPDT, LHAPDF and ROOT tools.
In addition, the \mosesWithSpace structure enables to form new independent modules that assemble the new model, as it was originally built to allow programming of complicated models with several non-standard aspects like the KK model.
In the new BSM models, we include new models where two initial state particles interact and create a maximum of three final state particles, {\em ie}, $2\to 1$, $2\to 2$ and $2\to 3$ hard-processes described by a differential cross-section function, as these are the constraints introduced by Pythia8.
Within \mosesNoSpace, these processes (cross-section functions) can be both analyzed independently and interfaced with Pythia8 to generate the corresponding fully simulated physics events.

Pythia8, the new generation of the commonly used event generators, Pythia, is a powerful and convenient tool.
Of the most important feature in this context, is the care that has been taken to allow user supplied hard subprocesses to be quickly and fully integrated into the Pythia8 framework.
This was the reason the \mosesWithSpace framework was developed on top of Pythia8, and will be enhanced with more interfaces and examples in the future.

It is suggested that introducing new processes will be forehanded by a validation procedure using analogous processes.
Since the presented KK implementation has many common characteristics with the corresponding SM implementation, a validation procedure was adopted.
The internal Pythia8 SM process was reproduced as a user process but while adjusting the helicity-amplitude formalism intended for the KK scenario.
The helicity amplitude was sent to Pythia8 using its standard \shellfont{SigmaProcess} interface class.
The $2 \to 2$ interface was used although it is also possible to utilize the $2\to 1$ production and subsequent $1\to 2$ decay.
The externally generated events were then compared with those from the self internal implementation subprocesses of Pythia8.
Throughout the following sections, the ``external / internal processes'' nomenclature will be used in this context.
Throughout the presented work the MRST2001lo parton distribution set~\cite{MRST} was used.

\section{Validation processes}
In the following section two methodical case studies are presented to validate the implementation of the external processes against Pythia8's internal processes; {\em(a)} the SM $e^+e^- \to \gamma \to \mu^+\mu^-$ at low energies where the contribution of the $Z^0$ boson is negligible and {\em (b)} the SM $q \bar q \to \gamma / Z^0 \to l^+ l^-$ at higher energies where the contribution of the $Z^0$ is dominant. 
Apart from the validation objective of these two examples, it is worthwhile going through these cases in some detail since it will also serve to rigorously explain the formalism used in the remainder of this paper.

\subsection[The SM $s$-channel photon exchange $e^+e^- \to \gamma \to \mu^+\mu^-$]{The SM $\boldsymbol{s}$-channel photon exchange $\boldsymbol{e^+e^- \to \gamma \to \mu^+\mu^-}$}
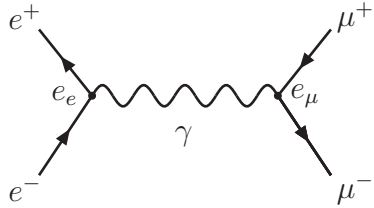
\begin{figure}[!th]
\begin{center}
\begin{picture}(300,70)(0,10) 
\SetWidth{1}
\Text(85,65)[c]{$e^+$}
\ArrowLine(110,35)(90,60) 
\Text(85,0)[c]{$e^-$}
\ArrowLine(90,5)(110,35) 
\Text(210,65)[c]{$\mu^+$}
\ArrowLine(200,60)(180,35) 
\Text(210,0)[c]{$\mu^-$}
\ArrowLine(180,35)(200,5) 
\Text(145,20)[c]{$\gamma$}
\Photon(110,35)(180,35){4}{4.5}
\ArrowLine(180,35)(200,5) 
\Text(100,35)[c]{$e_e$}
\Vertex(110,35){1.5}
\ArrowLine(180,35)(200,5) 
\Text(191,35)[c]{$e_\mu$}
\Vertex(180,35){1.5}
\end{picture}
\caption{\textsl{The $s$-channel photon exchange Feynman diagram}}
\label{fig:sChannelPhoton}
\end{center}
\end{figure}

In this case the Pythia8 output of the  estimated total cross section values using the external implementation and using Pythia8's internal scheme are compared.
The integrated total cross section $\sigma(s)$,
\begin{equation}
\sigma \left(s \right) = \frac{{4\pi \alpha_{em}^2 }}{3}\frac{{e_e^2 e_\mu^2 }}{s},
\label{eq:sigmaLEP_total}
\end{equation}
where $\hbar = c = 1$, can be realized by averaging the differential cross section over all the incoming helicity states, (corresponding to an unpolarized beam), by summing over all the outgoing helicity states, (corresponding to an unmeasured final polarization state) and by integrating over the solid angle $d\Omega$.
The differential cross section itself in terms of the helicity of the incoming electron $\lambda_{e^-}$ and the helicity of the outgoing muon $\lambda_{\mu^-}$ is
\begin{equation}
\frac{{d\sigma \left( {s,\cos \theta } \right)}}{{d\Omega }} = \frac{{\alpha_{em}^2 }}{{4s}}\frac{{s^2}}{{\left( {2S_{e^-} + 1} \right)\left(2S_{e^+} + 1 \right)}}\sum\limits_{\lambda_{e^-} = \pm \frac{1}{2}} {\sum\limits_{\lambda_{\mu^-}  = \pm \frac{1}{2}} {\left| {\frac{{e_e e_\mu }}{s}} \right|^2 \left( {1 + 4\lambda_{e^-} \lambda_{\mu^-} \cos \theta } \right)^2 } }
\label{eq:sigmaLEP_diffr}
\end{equation}
where, $\sqrt s$ is the collision's CM energy, the quantities $e_e$ $e_\mu$ are the charges (in units of the proton charge) of the leptons, see the corresponding tree-level diagram in Fig~\ref{fig:sChannelPhoton}.

The number of possible incoming helicity states $\left(2S_{e^+}+1\right)\left(2S_{e^-}+1\right)$ is expressed here in terms of the spins of the colliding particles.
From helicity conservation, it is sufficient to sum only over the helicity states of the incoming electron and the outgoing muon.
Finally, the polar angle $\theta$ is the angle of the outgoing $\mu^-$ relative to the incoming $e^-$ (with the azimuthal angle $\phi$, distributed uniformly). If one is to use the differential cross section from Eq~\ref{eq:sigmaLEP_diffr} in the \shellfont{SigmaProcess} $2 \to 2$ interface class of Pythia8, it is expected to be given in the Mandelstam variable $t$ related to $\cos \theta$ as
\begin{equation}
t = -\frac{s}{2}(1-\cos \theta).
\label{eq:mandelstam}
\end{equation}
The transformation introduces an extra $\frac{2}{s}$ factor in the cross section.
Note that by doing so, the differential cross section is now dimensionally different than the previous by $\frac{1}{s}$ and therefore has the units of GeV$^{-4}$, as required by Pythia8,
\begin{equation}
\frac{{d\sigma}}{{dt}} = \frac{2}{{s}}\frac{{d\sigma}}{{d\cos \theta}}.
\label{eq:transformation}
\end{equation}
After integrating over the azimuthal angle $\phi$, the correct function given to Pythia8 is
\begin{equation}
\frac{{d\sigma}}{{dt}} = \frac{2}{s} 2\pi \frac{{\alpha_{em}^2 }}{{4s}}\frac{s^2}{4}\sum\limits_{\lambda_e = \pm \frac{1}{2}} {\sum\limits_{\lambda_\mu = \pm \frac{1}{2}} {\left| {\frac{{e_e e_\mu }}{s}} \right|^2 \left( {1 + 4\lambda_e \lambda_\mu \cos \theta } \right)^2 }}
\label{eq:sigmaLEP_diffr_full}
\end{equation}

One can use Eq~\ref{eq:sigmaLEP_diffr_full} to generate events with low CM (CM) energy -- at 20 GeV, far enough below the $Z^0$ pole.
In this case it acceptable to ignore at the first approximation the $Z^0$ contribution.
All parton-level switches are turned off so the events are generated up to the level of the hard-process ({\em ie}, no parton showering, hadronization, kinematic cuts, etc.)

For comparison, the generation of the hard-subprocess is performed using the above but also repeated with the same run conditions and 
with the same sample size of 1M events but simply calling the internal process.
In fact, in Pythia8 this is usually performed in two steps; the $2 \to 1$ production and then the $1 \to 2$ decay.
However, the statistical error involved in the $2 \to 1$ production is very small since the angle is already integrated out and therefore, all phase space points will be the same and evaluate to the same value.
Therefore, the error which is subjected to roundoff errors, is unrealistic small and can be ignored.
For this reason, there's also a less familiar $2 \to 2$ implementation (in Pythia8) which was chosen for comparison since in this way one can get different cross sections depending on the angle selected on event by event basis and that leads to different event weights, which implies a realistic statistical error but also a lower efficiency.
The $2 \to 2$ Pythia8 internal implementation had to be slightly modified to fix the final state at $\mu^+\mu^-$.

The external function (see Eq~\ref{eq:sigmaLEP_diffr_full}) describes the same process and was implemented as a $2 \to 2$ with the intermediate photon included explicitly in it.
The results are summarized in Table~\ref{table:totalsigma_e+e-mu+mu-} where the agreement between these two values is to within $\sim$1-sigma.
\begin{table}[!th]
\centering
\caption{\textsl{Cross section statistics as given by Pythia8 for the two 1M samples (internal \& external) of $e^+e^- \to \gamma \to \mu^+\mu^-$ events at 20 GeV CM energy.}}
\vspace{4mm}
\begin{tabular}{cc}\hline\hline
Source code &  \hspace{10mm} Estimated $\sigma_{\rm{total}}$[nb]	\\
\hline
Internal	& \hspace{10mm} $0.2391 \pm 0.0001$	\\
External	& \hspace{10mm} $0.2389 \pm 0.0001$	\\
\hline\hline
\end{tabular}
\label{table:totalsigma_e+e-mu+mu-}
\end{table}


\subsection[The SM $s$-channel $\gamma/Z^0$ boson exchange $q \bar q \to \gamma / Z^0 \to l^+ l^-$]{The SM $\boldsymbol{s}$-channel $\boldsymbol{\gamma/Z^0}$ boson exchange $\boldsymbol{q \bar q \to \gamma / Z^0 \to l^+ l^-}$}
As with the previous example, this process can also provide useful information when validating our mechanism.
As before, all parton-level switches are turned off.
The first stage is to compare the estimated total cross section values of the external implementation with the internal implementation from Pythia8 for the hard process products only.
The second stage is the detailed comparison of the differential cross sections.

Since initially the focus was only on low masses, only the $\gamma$ contribution was taken into account.
At higher energies we must fully take into consideration the contributions from the $Z^0$ boson and $\gamma-Z^0$ interference terms.
By looking only at the leptonic final states which are produced by the $s$-channel processes we can ignore the more general $q \bar q \to f \bar f$ processes produced by both the $s$- and the $t$-channel exchange where $f$ can be any fermion.
This is experimentally useful, since the background from the SM QCD interactions were quark and gluon final states to be considered would be very high.

\subsubsection[Hadronic level: $pp \to \gamma \to l^+l^-X$]{Hadronic level: $\boldsymbol{pp \to \gamma \to l^+l^-X}$}
The Drell-Yan cross section for $q \bar q$ annihilation to a charged lepton pair via an intermediate massive photon can be easily obtained from the fundamental $e^+ e^- \to \gamma \to \mu^+ \mu^-$ cross section by the introduction of the appropriate color $N_C^q$ factors and by replacing the electron charge with that of the quark,
\begin{equation}
\sigma_{q\bar q \to \gamma \to l^+ l^-}\left(\hat s \right) = \hat \sigma_0 \frac{e_q^2 e_l^2}{N_C^q}
\label{eq:sigmaLHC_total}
\end{equation}
where $\hat \sigma_0 = \frac{4\pi \alpha_{em}^2}{3\hat s}$ and the overall color factor $\frac{1}{N_C^q}=\frac{1}{3}$ arises from to the fact that only when the color of the quark matches the color of the antiquark can annihilation into a color-singlet, leptonic, final state take place. 

The quantity $\hat s$ introduced here, as well as $\hat t$ which will be introduced below are the partonic Mandelstam variables while $s$ and $t$ are these defined for the incoming hadrons.
In general, the incoming quark and antiquark will have a spectrum of CM energies and it is more appropriate to consider the hadronic differential cross section $\frac{d\sigma}{d\hat s}$.
In order to obtain this, one can start from the CM frame of the 2 hadrons.
In this frame, the four momenta, $p_1^\mu$ and $p_2^\mu$ of the incoming partons may be written as 
\begin{eqnarray*}
p_1^\mu &=& \frac{\sqrt s}{2}\left( x_1 ,0,0,x_1 \right)\\
p_2^\mu &=& \frac{\sqrt s}{2}\left(x_2,0,0,-x_2\right).
\end{eqnarray*}
The square of the parton CM energy $\hat s$ is related to the corresponding hadronic quantity by $\hat s = x_1 x_2 s$.
Folding in the parton distribution functions for the initial state quarks and antiquarks gives the hadronic differential cross section in terms of $x_1$ and $x_2$,
\begin{equation}
\frac{d^2\sigma}{dx_1 dx_2} = \frac{\hat \sigma_0(\hat s)}{N_C^q} e_l^2 \sum\limits_q{e_q^2 \left[\mathcal{F}_q(x_1 ,Q) \mathcal{F}_{\bar q}(x_2,Q) + \left\{ {1 \leftrightarrow 2} \right\} \right]}
\label{eq:sigmaLHCpdf}
\end{equation}
where $\mathcal{F}_q$ is the parton density function of species $q$.
The quantity $Q$ is the factorization scale, usually taken to be the invariant mass, $\sqrt{\hat s}$.
From beam symmetry, the substitution $\left\{ {1 \leftrightarrow 2} \right\}$ is equivalent to simply multiplying the whole expression by $2$.
To obtain the hadronic differential cross section the transformation from $x_1$ and $x_2$ to $\hat s$ and $y$ is necessary,
\begin{equation}
\begin{array}{rl}
	&\hat s = s x_1 x_2 \\
	&y = \frac{1}{2}\ln{(\frac{x_1}{x_2})}.
\end{array}
\label{eq:rapidity}
\end{equation}
where $y$ is the rapidity of the pair.
The transformation given in Eq~\ref{eq:rapidity} involves a Jacobian which reduces to the constant $\frac{1}{s}$.
It is also necessary to integrate over all possible rapidity values, where this integration is usually done numerically.
By knowing the value of $s$ (14 TeV for the LHC), the hadronic differential cross section can written as
\begin{equation}
	\frac{d\sigma}{d\hat s} = \frac{\hat \sigma _0}{N_C^q} e_l^2 \int\limits_{-y_0}^{+y_0} \frac{dy}{s} \sum\limits_q{e_q^2 \left[\mathcal{F}_q\left(x_1(y,\hat s) ,\sqrt{\hat s}\right) \mathcal{F}_{\bar q}\left(x_2(y, \hat s), \sqrt{\hat s}\right) + \left\{ {1 \leftrightarrow 2} \right\} \right]}
\label{eq:sigmaLHCpdf_y}
\end{equation}
where the boundaries $\pm y_0$ are determined from both $x_1$ and $x_2$ being constrained between $0$ and $1$ so that 
$y_0 = \frac{1}{2}\ln\left(\frac{s}{\hat s}\right)$.
It is sometimes useful to replace the transformation Jacobian with the equivalent expression, $\frac{1}{s} = \frac{x_1 x_2}{\hat s}$.

\subsubsection[Partonic level: $q \bar q \to \gamma \to l^+ l^-$]{Partonic level: $\boldsymbol{q \bar q \to \gamma \to l^+ l^-}$}
The differential cross-section function which Pythia8 takes should describe the hard-process itself.
It should not be given as the (hadronic) differential cross section and it should not include any parton density functions.
This is because $\hat s$ is being determined separately for every generated event and since the evolution of the parton distributions is performed internally by Pythia8.
This is similar to the $e^+e^- \to \gamma \to \mu^+\mu^-$ process considered previously where only the hard process was considered. 
Inserting the $\hat t$ dependency, the function given to Pythia8 becomes
\begin{equation}
\frac{d\hat \sigma \left( {\hat s,\cos \theta^* } \right)}{d\hat t} = \frac{2}{\hat s} 2\pi \frac{\alpha_{em}^2}{4\hat s}\frac{1}{N_C^q}\frac{\hat s^2}{4}\sum\limits_{\lambda_q = \pm \frac{1}{2}} {\sum\limits_{\lambda_l = \pm \frac{1}{2}} {\left| {\frac{e_q e_l}{\hat s}} \right|^2 \left( {1 + 4\lambda_q \lambda_l \cos \theta^* } \right)^2 } }
\label{eq:sigmaLHC_ds_for_Pythia}
\end{equation}
where this is written in the CM frame of the incoming partons.
The angle $\theta^*$ is the polar angle in the CM frame between the incoming $q$ and the outgoing $l^-$ in contrast to the polar angle, $\theta$, in the lab frame.

\subsubsection[A $q\bar q$ final state]{A $\boldsymbol{q\bar q}$ final state}
\begin{figure}[!th]
\begin{center}
\begin{picture}(480,70)(0,0) 
\SetWidth{1}

\Text(5,65)[c]{$\bar q_{\bar c}$}
\ArrowLine(30,35)(10,60) 
\Text(5,0)[c]{$q_c$}
\ArrowLine(10,5)(30,35) 
\Text(125,65)[c]{$\bar q_{\bar c'}$}
\ArrowLine(120,60)(100,35) 
\Text(125,0)[c]{$q_{c'}$}
\ArrowLine(100,35)(120,5) 
\Photon(30,35)(100,35){4}{6.5}
\Vertex(30,35){1.5}
\Vertex(100,35){1.5}
\Text(65,20)[c]{$\gamma$}

\Text(135,35)[c]{+}

\Text(145,65)[c]{$\bar q_{\bar c}$}
\ArrowLine(185,55)(150,65) 
\Text(145,0)[c]{$q_{c'}$}
\ArrowLine(150,5)(185,15) 
\Text(225,65)[c]{$\bar q_{\bar c}$}
\ArrowLine(220,65)(185,55) 
\Text(225,0)[c]{$q_{c'}$}
\ArrowLine(185,15)(220,5) 
\Photon(185,55)(185,15){3}{4.5}
\Vertex(185,55){1.5}
\Vertex(185,15){1.5}
\Text(165,35)[c]{$\gamma$}

\Text(233,35)[c]{+}

\Text(243,65)[c]{$\bar q_{\bar c}$}
\ArrowLine(268,35)(248,60) 
\Text(243,0)[c]{$q_c$}
\ArrowLine(248,5)(268,35) 
\Text(363,65)[c]{$\bar q_{\bar c'}$}
\ArrowLine(358,60)(338,35) 
\Text(363,0)[c]{$q_{c'}$}
\ArrowLine(338,35)(358,5) 
\Gluon(268,35)(338,35){4}{6.5}
\Vertex(268,35){1.5}
\Vertex(338,35){1.5}
\Text(303,20)[c]{$g$}

\Text(370,35)[c]{+}

\Text(381,65)[c]{$\bar q_{\bar c}$}
\ArrowLine(421,55)(384,65) 
\Text(381,0)[c]{$q_{c'}$}
\ArrowLine(384,5)(421,15) 
\Text(461,65)[c]{$\bar q_{\bar c}$}
\ArrowLine(454,65)(421,55) 
\Text(461,0)[c]{$q_{c'}$}
\ArrowLine(421,15)(454,5) 
\Gluon(421,55)(421,15){3}{4.5}
\Vertex(421,55){1.5}
\Vertex(421,15){1.5}
\Text(410,35)[c]{$g$}

\end{picture}
\caption{\textsl{The competing amplitudes for a production of a di-quark final state. The notations $c,c',\bar c, \bar c'$ account for the correct color flow. The leading contributions come from the gluons exchange.}}
\label{fig:s_t_Channel}
\end{center}
\end{figure}
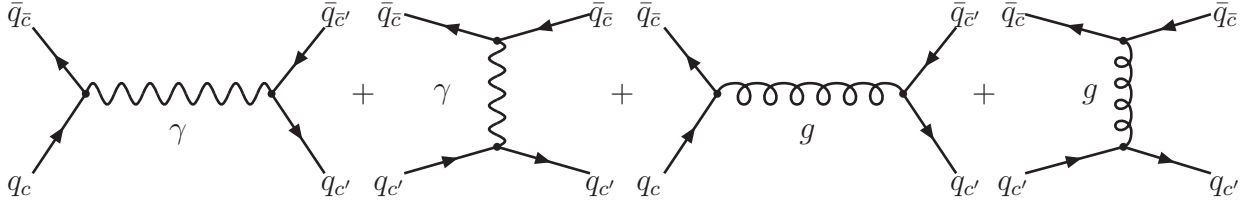

So far account has only been taken for the $s$-channel $\gamma$ exchange, since only the leptonic final state, different from the initial state, at only low energies has been considered.
To generalize this Drell-Yan annihilation to any pair of fermions, $q\bar q \to \gamma \to f\bar f$, Eq~\ref{eq:sigmaLHC_ds_for_Pythia} must be modified.
The cross section has to be multiplied by the appropriate color factor $N_C^f$ and all the existing lepton indices must be substituted with corresponding fermion indices $l \leftrightarrow f$.
The new overall color factor is $N_C^f=3$ for $f = q$ or $N_C^f=1$ for $f = l$.
Doing so, it is apparent that in the general di-fermion final state the contribution of the $t$-channel exchange should also be considered. This is since the di-fermion final state can consist of the same (annihilated) quark-antiquark pair, $q \bar q \to f \bar f =\left(q \bar q\right)_{\rm{same}}$ and the exchange can take place in either of the $s$-channel or the $t$-channel either with photon or gluon exchange as can be seen in Fig~\ref{fig:s_t_Channel}.
Thus, a di-jet final state can also be observed.
Indeed, the cross section should be much larger than for the di-lepton final state.
However, the contribution of the $s$-channel $\gamma$ exchange to this di-jet final state is negligible since the electroweak interaction 
is much weaker than the strong interaction.
Therefore, this channel will be dominated by the exchange of a gluon, either in the $s$- or the $t$-channel.
The $t$-channel exchange is expected to be significantly larger than the corresponding $s$-channel exchange, in particular at forward angles.
However, experimentally it depends on the transverse momentum cut, $p_T$, applied on the outgoing quark, or jet, hadronic state.
The $t$-channel exchange predominantly produces events with smaller $p_T$.
In addition, there are significantly more QCD diagrams involving gluons in the initial and final states which can lead to di-jet production.
In practice, it is not possible to distinguish between the initial state and the final state sources.
It is very difficult to distinguish between a quark and gluon jet, unless for instance, it can be identified as a $b$-jet by 
the presence of a displaced vertex.
In principle it may be possible to suppress the $t$-channel contributions by a rejection based on the di-jet system $p_T$.
Nevertheless, for simplicity, the remainder of this paper will address only di-lepton production from $s$-channel exchange.

\subsubsection[Including the $Z^0$ boson: $q \bar q \to \gamma/Z^0 \to l^+l^-$ at $\sqrt{\hat s} \geq m_{Z^0}$]{Including the \boldmath$Z^0$ boson: \boldmath$q \bar q \to \gamma/Z^0 \to l^+l^-$ at \boldmath$\sqrt{\hat s} \geq m_{Z^0}$}
With increasing energies, around and above the $Z^0$ pole, the $s$-channel $Z^0$ intermediate state and interference terms must be included.
The helicity amplitude of the $Z^0$ boson has to be added to the photon amplitude, $e_q e_l/\hat s$, to account for the correct interference effects,
\begin{equation}
	\tilde M_{\lambda_q \lambda_l}\left(\hat s \right) = \frac{e_q e_l}{\hat s} + \frac{{g_{\lambda_q } g_{\lambda_l} }}{{\hat s - m_{Z^0}^2 + i \left(\frac{\hat s}{m_{Z^0}}\right) \sum\limits_F\Gamma_{Z^0 \to F \bar F}}}.
	\label{eq:totalamplitude}
\end{equation}
Here, the SM coupling constants~\cite{ZLINESHAPE} of the $Z^0$ gauge boson to the involved fermions are,
\begin{equation}
		g_{\lambda_f} =
			\left\{
				\begin{array}{rl}
					-\frac{ e_f \sin^2 \theta_W}{\sin \theta_W \cos \theta_W} &\mbox{if $\lambda_f=+1/2$} \\\\
					\frac{I_f^3 - e_f \sin^2 \theta_W}{\sin \theta_W \cos \theta_W} &\mbox{if $\lambda_f=-1/2$}
				\end{array} \right.
	\label{eq:couplings}
\end{equation}
where the index $f$ represents either the incoming quark or the outgoing lepton.
Here the additional values for the $Z^0$ boson mass, $m_{Z^0}$, the weak isospin of the incoming quarks or the outgoing leptons, $I_f^3$ and the weak-mixing angle, $\theta_W$ have been introduced.
The sum $\sum\limits_F\Gamma_{{Z^0} \to F\bar F}$ is the total $Z^0$ decay width to all fermion-antifermion pairs denoted by $F \bar F$, where the partial decay width in Lowest Order (LO) is:
\begin{equation}
\Gamma_{Z^0 \to F\bar F} = \frac{N_C^F \alpha_{em} m_{Z^0}}{6} \left( \left| g_{+\frac{1}{2}} \right|^2 + \left| g_{-\frac{1}{2}} \right|^2 \right).
\label{eq:width}
\end{equation}
To account for radiative corrections, the relation $\frac{\pi \alpha_{em}}{\sqrt{2}G_{\mu}} = m_{Z^0}^2\sin^2\theta_W\cos^2\theta_W$ has been used, where $G_\mu$ is the muon decay constant and $m_W$ is the $W$ boson mass.
The corrected expression for the partial decay width is
\begin{equation}
\Gamma_{Z^0 \to F\bar F} = \frac{G_{\mu} N_C^F m_{Z^0}^3}{3 \pi \sqrt{2}} \left[ \left(I_F^3\right)^2 - 2 I_F^3 e_F \sin^2 \theta_W + 2\left(e_F \sin^2 \theta_W\right)^2 \right].
\label{eq:width_radiative_corrections}
\end{equation}
Finally, the differential cross section in its LO takes the form
\begin{equation}
	\frac{d\hat \sigma \left(\hat s,\cos \theta^* \right)}{d\hat t} = \frac{2}{\hat s} 2\pi \frac{\alpha_{em}^2}{4\hat s}\frac{1}{N_C^q}\frac{\hat s^2}{4}\sum\limits_{\lambda_q = \pm \frac{1}{2}} {\sum\limits_{\lambda_l = \pm \frac{1}{2}} {\left| \tilde M_{\lambda_q \lambda_l}\left(\hat s \right) \right|^2 \left( {1 + 4\lambda_q \lambda_l \cos \theta^* } \right)^2 } }.
	\label{eq:sigmatotalamplitude}
\end{equation}
In this context it is worth mentioning that self-formulated coupling constants were used in Eq~\ref{eq:couplings}.
These can be transformed into the Pythia8 coupling constants, $a_f, v_f, L_f, R_f$ given by the \shellfont{CoupEW} class. 
The couplings used in \mosesWithSpace can be modified without causing undesired behavior in Pythia8.
Thus, they provide secure flexibility to probe non-SM phenomena that depend on them.
The same considerations are also relevant for the widths.

\subsubsection{The angular decay asymmetry}
To expand the previous discussion it is useful to define two additional reference frames:
{\em (a)} the colliding proton CM frame denoted by $\mathcal{O}$ (this frame is identical to the laboratory frame) and, 
{\em (b)} the rest frame of the di-lepton system denoted by $\mathcal{O}^*$.
Neglecting higher order processes, the di-lepton system is, in general boosted along the beam axis.
The $z$-axis is arbitrarily chosen as the direction of one of the beams, and it is then identical for $\mathcal{O}$ and $\mathcal{O}^*$ frames.
One of the primary observables for this process is the forward-backward asymmetry that can be extracted from the distribution of $\cos{\theta^*}$ in the $\mathcal{O}^*$ frame.
By definition, it is the cosine of the angle between the quark and the lepton directions in the $\mathcal{O}^*$ frame.
It should be noted that there is a sign ambiguity in the measurement of $\cos{\theta^*}$, since for a particular event, there is no information about whether the incoming quark comes from the positive or negative  $z$ directions.
Instead, it is useful to consider the quantity $\cos{\theta_\beta^*}$, where $\theta_{\beta}^*$ is the angle between the di-lepton system boost $\vec{\beta}$ (relative to the $\mathcal{O}$ frame) and the lepton direction
\begin{equation}
	\cos{\theta_{\beta}^*} = \frac{\vec{p^*}_{l} \cdot \vec{\beta}}{\left|\vec{p^*}_{l}\right| \cdot |\vec{\beta}|}
\label{eq:costhetabeta}
\end{equation}
where the boost vector is $\vec{\beta} = \frac{\vec{p}_{l} + \vec{p}_{\bar l}}{E_{l} + E_{\bar l}}$.
In order to obtain $\vec{p^*}_{l}$, the boost vector of the di-lepton system should be found and the transformation to the $\mathcal{O}^*$ frame should be performed.
Neglecting higher order processes, the boost is confined to the $\left|z\right|$ direction, that is, $\vec{\beta} = \beta \hat z$, and can be measured so that there is no sign ambiguity in determining $\cos\theta^*_{\beta}$.

The next step is to calculate the forward-backward asymmetry $A_{fb}$.
Here we define
\begin{equation}
	A_{fb}\left(\hat s\right) = \frac{d\sigma_{f}/d\hat s - d\sigma_{b}/d\hat s}{d\sigma_{f}/d\hat s + d\sigma_{b}/d\hat s} = \frac{d\sigma_{f}/d\hat s - d\sigma_{b}/d\hat s}{d\sigma_{\rm{tot}}/d\hat s}
\label{eq:Afb_theory}
\end{equation}
where, in the absence of any detector cuts applied, the quantity $d\sigma_{f/b}/d\hat s$ is given by
\begin{equation}
\frac{d\sigma_{b}}{d\hat s} = \int\limits_{-1}^0{\frac{d\sigma\left(\hat s\right)}{d\hat s d\cos\theta^*} d\cos\theta^*} \,\,\,\,\,\,\,\,\,\,\,\,\,\,\,\,\,\,
\frac{d\sigma_{f}}{d\hat s} = \int\limits_0^1{\frac{d\sigma\left(\hat s\right)}{d\hat s d\cos\theta^*}d\cos\theta^*}.
\label{eq:sigma_fb}
\end{equation}
Since the integrand is given in terms of $\cos\theta^*$ it should be rewritten to enable re-classification of the forward and backward definitions.
This can be done by separating into different rapidity contributions since $y$ and $\beta$ have the same sign,
\begin{equation}
\begin{array}{rl}
&\frac{d\sigma_{b}^{\beta}}{d\hat s} = \int\limits_{-y_0}^{0}dy \int\limits_{0}^1{\frac{d\sigma\left(\hat s, y\right)}{d\hat s d\cos\theta^*} d\cos\theta^*} + \int\limits_{0}^{+y_0}dy \int\limits_{-1}^0{\frac{d\sigma\left(\hat s, y\right)}{d\hat s d\cos\theta^*} d\cos\theta^*}\\\\
&\frac{d\sigma_{f}^{\beta}}{d\hat s} = \int\limits_{-y_0}^{0}dy \int\limits_{-1}^0{\frac{d\sigma\left(\hat s, y\right)}{d\hat s d\cos\theta^*} d\cos\theta^*} + \int\limits_{0}^{+y_0}dy \int\limits_{0}^1{\frac{d\sigma\left(\hat s, y\right)}{d\hat s d\cos\theta^*} d\cos\theta^*}
\end{array}
\label{eq:sigma_fb_y}
\end{equation}
where $y_0$ is the rapidity kinematic limit (see Eq~\ref{eq:sigmaLHCpdf_y}) and the form of Eq~\ref{eq:Afb_theory} remains the same under the substitutions $\sigma_{f/b} \to \sigma_{f/b}^{\beta}$ and $A_{fb} \to A_{fb}^{\beta}$.
The integration over $y$ in Eq~\ref{eq:sigma_fb_y} is performed asymmetrically and thus, the relation in Eq~\ref{eq:sigmaLHCpdf} cannot be used.
Therefore, in the hadronic level the differential cross section is usually separated into two terms with respect to $\cos\theta^*$: Symmetric and Anti-symmetric denoted by $S$ and $A$.
This enables the separate contributions that form the forward-backward asymmetry to be identified, 
\begin{equation}
\frac{d\sigma}{dy d\hat s d\cos\theta^*} \sim {\hat s^2}\sum\limits_q{\left[G_q^S\left(y,\hat s\right)S_q\left(\hat s\right)\left(1+ {\cos^2\theta^*}\right) + G_q^A\left(y,\hat s\right)A_q\left(\hat s\right) 2\cos\theta^* \right]}
\label{eq:sigmaLHCpdf_y_SandA}
\end{equation}
where the Symmetric and Anti-symmetric combinations $G_{\bar q}^{S/A=+/-}$ that involve the parton density functions of the colliding hadrons are defined
\begin{equation}
	G_{q}^{S/A} = \frac{x_1 x_2}{\hat s} \left[\mathcal{F}_q\left(x_1 ,\sqrt{\hat s}\right) \mathcal{F}_{\bar q}\left(x_2, \sqrt{\hat s}\right) \pm  \mathcal{F}_{\bar q}\left(x_1 ,\sqrt{\hat s}\right) \mathcal{F}_q\left(x_2, \sqrt{\hat s}\right)\right]
\label{eq:pfd_SA}
\end{equation}
where $x_1$ and $x_2$ can be written in terms of $y$ and $\hat s$ so $G_q$ is in fact a function of $y$ and $\hat s$.
The terms $S_q$ and $A_q$ can be realized from the helicity amplitude:
\begin{equation}
\begin{array}{rl}
&S_q\left(\hat s\right) = \sum\limits_{\lambda_q} {\sum\limits_{\lambda_l} {\left| \tilde M_{\lambda_q \lambda_l} \right|^2}} \\
&A_q\left(\hat s\right) = \sum\limits_{\lambda_q = \lambda_l} {\left| \tilde M_{\lambda_q \lambda_l} \right|^2} - \sum\limits_{\lambda_q \neq \lambda_l} {\left| \tilde M_{\lambda_q \lambda_l} \right|^2}
\end{array}
\label{eq:SqAq}
\end{equation}
where $\tilde M_{\lambda_q \lambda_l}$ from Eq~\ref{eq:totalamplitude} is sensitive to the couplings of the $Z^0$ boson to the different families of SM fermions.
Using Equations~\ref{eq:sigmaLHCpdf_y_SandA},~\ref{eq:pfd_SA} and~\ref{eq:SqAq}, it can be seen that under these definitions the forward backward-asymmetry expression reduces to
\begin{equation}
	A_{fb}^{\beta}\left(\hat s\right) = \frac{\sum\limits_q A_q\left(\hat s\right) \int\limits_0^{+y_0}dy G_{q}^{A}\left(y,\hat s\right) - \sum\limits_q A_q\left(\hat s\right) \int\limits_{-y_0}^0dy G_{q}^{A}\left(y,\hat s\right)}{\sum\limits_q S_q\left(\hat s\right) \int\limits_{-y_0}^{+y_0}dy G_{q}^{S}\left(y,\hat s\right)}.
\label{eq:Afb_pdf}
\end{equation}
In cases where the integration over $\cos\theta^*$ is performed in a smaller effective interval ($\left[-k,k\right]$ where $0 < k < 1$) due to some kinematic cuts, then the substitutions $\int dy \to \int dy k^2$ and $\int dy \to \frac{3}{4}\int dy \left(k+\frac{k^3}{3}\right)$ can be made in the numerator and the denominator respectively.
These substitutions will be relevant in the following discussion where the $ATLAS$ detector cuts are introduced.

Thus, the origin of the asymmetry is understood and can be summarized by simply considering the $\cos\theta^*$ distribution
\begin{equation}
	\frac{1}{N}\frac{dN}{d\cos\theta^*_{\beta}} = \frac{3}{8} \left[ 1 + \cos^2\theta^*_{\beta} \right] + A_{fb}^{\beta}\cos\theta^*_{\beta}. 
\label{eq:costhetadistribution}
\end{equation}
The forward-backward asymmetry coefficient $A_{fb}^{\beta}$ depends on the couplings of the fermions to the $Z^0$ boson and is thus, sensitive to $\sin^2\theta_W$ and it can be extracted by fitting Eq.~\ref{eq:costhetadistribution} to the data. 
However, in the case of high statistics and where there are no kinematic cuts, a direct measurement could be done by simply counting the forward and backward events $N_{f/b}^{\beta}$.
These are manifestly given relative to the boost direction if $\cos\theta_{\beta}^*$ was first used
\begin{equation}
	A_{fb}^{\beta} = \frac{N_f^{\beta} - N_b^{\beta}}{N_f^{\beta} + N_b^{\beta}}.
\label{eq:Afb_analysis}
\end{equation}
A detailed discussion of the polar angle distribution and the asymmetry is given in the following 
sections including the effects of Initial State Radiation (ISR).

\subsubsection[Validation results with $\gamma/Z^0$]{Validation results with \boldmath$\gamma/Z^0$}
Using the expression given in Eq~\ref{eq:sigmatotalamplitude} events were generated simulating colliding proton beams at the CM energy of 14 TeV with no cuts applied, and all the parton-level switches were turned off.
Considering only the di-muon final state, all other decay modes of the $\gamma/Z^0$ were also turned off.
The same validation procedure described previously is repeated, combined with the comparison of the hadronic cross section shapes.
As a second validation stage, the code Pythia8 uses internally was copied and was plugged-in as if it was an external process.
In the following this is referred to as the semi-external validation.

\begin{figure}[!th]
	\centering
	\begin{overpic}[scale=0.85]{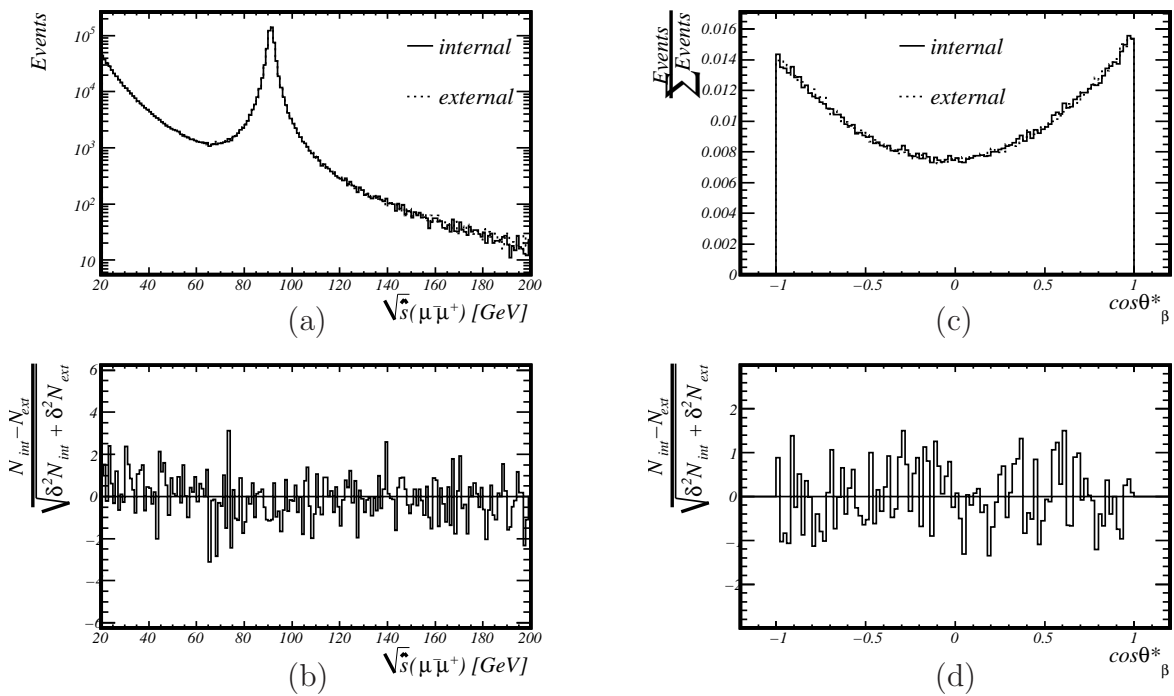}
		\put(110,135){(a)}
		\put(110,0){(b)}
		\put(355,135){(c)}
		\put(355,0){(d)}
	\end{overpic}
	\caption{\textsl{The results obtained with \mosesWithSpace \textbf{external} code (solid) and Pythia8 internal code (dashed). The $\sqrt{\hat s}$ distributions and their comparison in {\em (a)} and {\em (b)} and, the normalized $\cos\theta^*_{\beta}$ distributions around the $Z^0$ peak and their comparison in {\em(c)} and {\em(d)}.}}
	\label{fig:internalVSexternal}
\end{figure}

\begin{figure}[!th]
	\centering
	\begin{overpic}[scale=0.85]{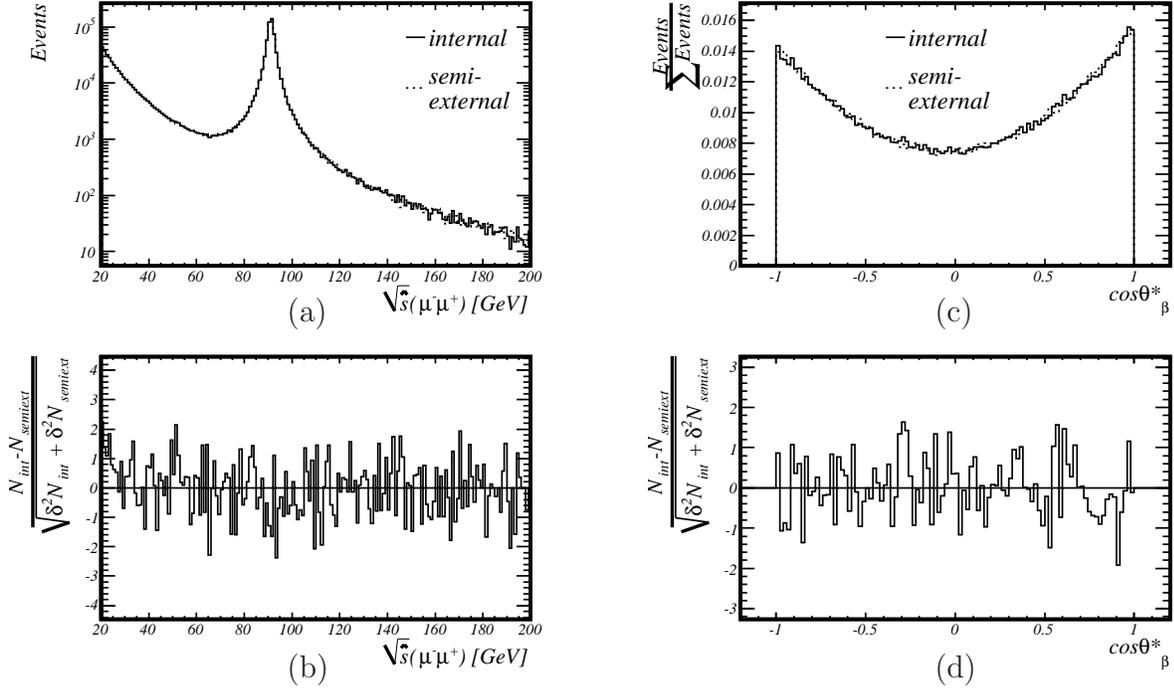}
		\put(110,135){(a)}
		\put(110,0){(b)}
		\put(355,135){(c)}
		\put(355,0){(d)}
	\end{overpic}
	\caption{\textsl{The results obtained with \mosesWithSpace \textbf{semi-external} code (solid) and Pythia8 internal code (dashed). 
The distributions description is identical to Fig~\ref{fig:internalVSexternal}.}}
	\label{fig:internalVSsemiexternal}
\end{figure}

The differential cross section shapes and the distributions of the $\cos\theta^*_{\vec{\beta}}$ are shown in Figures~\ref{fig:internalVSexternal}{\em a,c} and~\ref{fig:internalVSsemiexternal}{\em a,c} for the external and semi-external process implementations respectively.
Corresponding bin-by-bin comparisons are shown in Figures~\ref{fig:internalVSexternal}{\em b,d} and~\ref{fig:internalVSsemiexternal}{\em b,d}.
The numerical results are summarized in Table~\ref{table:totalsigma_qqbartoffbar}.
Note that since in this section, the generation is stopped at the hard process level, the asymmetry results are likely to be slightly 
modified by initial state radiation and will be discussed in a later section, but this should not, in practice, modify the agreement 
between the results of the three implementations.
\begin{table}[!th]
\centering
	\caption{\textsl{Numerical comparisons. The cross section statistics as given by Pythia8 and the $A_{fb}^{\beta}$ value around the $Z^0$ resonance. This is the output for the three 1M samples of $pp \to \gamma/Z^0 \to \mu^+ \mu^-$ events with 14 TeV CM energy. The lower mass cut-off is at 20~GeV.}}
\vspace{4mm}
	\begin{tabular}{cccc}\hline\hline
	Source code	& \hspace{0mm} Estimated $\sigma_{\rm{total}}$[nb]	& \hspace{0mm} $A_{fb}^{\beta}$	& \hspace{0mm} Events in $\left|\sqrt{\hat s} - m_{Z^0}\right| \leq \Gamma_{Z^0}^{\rm{tot}}$ \\
	\hline
	Internal			& \hspace{0mm} $2.624 \pm 0.001$		& \hspace{0mm} $0.0445 \pm 0.0018$	& \hspace{0mm} 308160\\
	Semi-External	& \hspace{0mm} $2.618 \pm 0.001$		& \hspace{0mm} $0.0459 \pm 0.0018$	& \hspace{0mm} 309381\\
	External			& \hspace{0mm} $2.619 \pm 0.001$		& \hspace{0mm} $0.0469 \pm 0.0018$	& \hspace{0mm} 309186\\
	\hline\hline
	\end{tabular}
	\label{table:totalsigma_qqbartoffbar}
\end{table}

The agreement between the total cross section values in Table~\ref{table:totalsigma_qqbartoffbar} is to within $\sim$3-sigma, where most of this difference might be due to the phase space sampling.
The forward-backward asymmetry values in Table~\ref{table:totalsigma_qqbartoffbar}, extracted from the three $\cos\theta_{\beta}^*$ distributions, agree to within $\sim$0.1-sigma.
Figures~\ref{fig:internalVSexternal}{\em b,d} and~\ref{fig:internalVSsemiexternal}{\em b,d} illustrate the agreement of differential distributions to within the respective statistical uncertainties.
A $\chi^2$ comparison  between the internal and external histograms yields $\frac{\chi^2}{\rm{DOF}} = 1.04$ for the di-lepton mass distribution and $\frac{\chi^2}{\rm{DOF}} = 1.03$ for the $\cos\theta_{\beta}^*$ distribution.
Likewise, the $\chi^2$ comparison test between the internal and semi-external histograms yielded $\frac{\chi^2}{\rm{DOF}} = 0.94$ for the di-lepton mass distribution and $\frac{\chi^2}{\rm{DOF}} = 1.12$ for the $\cos\theta_{\beta}^*$ distribution.

\clearpage

\section{The new KK boson exchange process}
The Drell-Yan KK process,  $pp \to \gamma^* / Z^* \to l^+ l^-X$, is similar to those discussed in the previous section and has not yet been implemented as an intrinsic component of any public Monte Carlo generator.
After following the validation procedure for the formalism above, it can be used to introduce the new specific KK process within the theoretical framework given below.
Also given below is a brief discussion on other possible heavy gauge bosons that can be produced in the LHC and share the same final di-lepton state.

\subsection{An overview of the theoretical framework}
\subsubsection{The Kaluza-Klein model}
The observable world consists of at least the three known spatial dimensions, and one of time, but it is possible that there are additional spatial dimensions that are not directly observable.
When considering these potential extra dimensions, it is useful to distinguish between parallel and transverse dimensions with respect to the 3$d$ world.
The size of all parallel dimensions should be constrained to be no larger than $\sim$ TeV$^{-1}$ ($10^{-18}$ m) in order to be unobservable at present energies.
The size of transverse dimensions remain unrestricted with much weaker experimental bounds.
One popular possibility~\cite{ANTONIADIS,ARKANIHAMED,ARKANIHAMED2,RIZZONEWLEPTONCOLLIDERS} is one extra parallel dimension, compactified on the $S^{1}/Z_{2}$ orbifold where gauge fields have KK excitations in this extra parallel dimension but where fermions are localized on the 3$d$ brane and have no KK excitations.
KK states of gauge bosons are then singly produced as new resonances that may be observed experimentally.
This description leads to equally spaced KK states of gauge fields with masses given by
\begin{equation}
m_n^2 = m_0^2 + \frac{n^2}{R^2}
\label{eq:kk_mass}
\end{equation}
where $m_n$ is the higher dimensional mass and $R$ is the radius of compactification.
The mode $n=0$ is identified with the $4d$ (SM) state, while the higher modes have the same quantum numbers as the lowest, but with increasing mass.
Within this model, the couplings to fermions of these excited modes are larger than the known couplings of the zero modes (SM bosons) by a factor of $\sqrt{2}$ due to the normalization of the KK excitations~\cite{ARKANIHAMED2,KKEXPERIMENTAL,KKCONSTRAINTS}.
In this paper, only in the excitations of the $Z^0$ and the photon shall be considered.
If such KK states exist, they have to exceed the lower bounds on their mass, based on their indirect effects associated with their tower exchange.
These bounds rely upon a number of additional assumptions, notably, that the effect of KK exchange is the only new physics beyond the SM.
In the 5$d$ case, a global fit to the precision electroweak data including the contributions from KK gauge interactions yields $R^{-1}\gtrsim$4~TeV~\cite{KKEXPERIMENTAL,KKCONSTRAINTS,PDG,KKEWGLOBALFIT}.
The reader is referred to the appendix for a derivation of the KK tower for a $5d$ real massless scalar field.
In the following discussion, the notation $m^* \equiv R^{-1}$ is used.

\subsubsection[Additional heavy $Z$-like boson production]{Additional heavy $\boldsymbol{Z}$-like boson production}
A similar deviation from the SM cross section coming from KK excitations of $\gamma/Z^0$, may be observed also in models with an additional heavy gauge boson~\cite{ZPRIMEPHENOMENOLOGY,ZPRIMEPHYSICS,ZPRIMETOE+E-}.
Several Grand Unified Theories (GUTs) postulate that the $SU(3)$, $SU(2)$ and $U(1)$ symmetry groups of the SM have a common origin as sub-groups of some larger symmetry group $G$.
It is supposed that at large energy scales, this symmetry is valid but below some critical energy scale, $G$, it is spontaneously broken.
This kind of GUT predicts at least one additional gauge boson after the symmetry is broken to the SM.
In these models, the extra neutral gauge bosons are usually denoted by $Z'$.

For comparison with the KK model, one specific model that is featuring at least one $Z'$ boson (light enough to be detected at the LHC) was considered.
These models can come from the breaking of the $E_6$ group which is a popular candidate to GUT symmetry.
The $Z'$ can be observed as a peak in the di-lepton mass distribution above a small background, and the LHC discovery potential for that is reasonably high and well known~\cite{ZPRIMESTUDIESINLHC}.
If a resonance is observed at the $Z'$ or $\gamma^*/Z^*$ hypothetical mass at the LHC, a discrimination mechanism between these two candidates would be required.
In specific circumstances, this discrimination is possible~\cite{ZPRIMETOE+E-,RIZZOASYMMETRY}, however, for both the generation of the KK events and the discrimination mechanism, a somewhat different approach is considered in this paper.
Out of the possible ways to break the $E_6$ group, the following shall be considered 
{\em (i)} $Z'_{\psi}: E_{6}\rightarrow SO(10)\times U(1)_{\psi}$, 
{\em (ii)}, $Z'_{\chi}: E_{6}\rightarrow SO(10)\times U(1)_{\psi}\rightarrow SU(5)\times U(1)_{\chi}\times U(1)_{\psi}$ and 
{\em (iii)} $Z'_{\eta}: E_{6}\rightarrow SM\times U(1)_{\eta}$.
The different couplings of the new gauge field to the SM fermions within these models can be found elsewhere~\cite{ZPRIMETOE+E-}.
Another scenario often introduced is the $Z'_{\rm{SM}}$ where the new boson has the same couplings as the $Z^0$ but with different mass and width.
One should note that there is no theoretical justification for the choice of SM-like couplings for the $Z'$.
However, this is the more experimentally challenging case since if there will be a resonance in the KK mass, then the case where the couplings of the $Z'$ are SM-like is practically the most difficult to distinguish from the KK resonance.
Therefore, this will be the choice for the following discussion, where this specific model shall be denoted by $Z'_{{\rm{SM}}}$.
In this paper, the mass of these new bosons are taken to be 4~TeV, same as the mass of the KK bosons.

\subsection{The KK implementation in the LHC scenario}
As mentioned in the introduction, the LHC represents a new frontier in the search for heavy resonances.
Each of the resonances in the KK tower discussed here could be produced by a similar mechanism to the light SM bosons.
From an experimental stand point, most of the considerations described in the previous sections hold also for the KK case.

In the parton level, the process $q \bar q  \to \gamma^* / Z^* \to l^+ l^-$ can be expressed in terms of the following differential cross section
\begin{equation}
	\frac{d\hat \sigma \left(\hat s,\cos \theta^* \right)}{d\hat t} = \frac{2}{\hat s} 2\pi \frac{\alpha_{em}^2}{4\hat s}\frac{1}{N_C^q}\frac{\hat s^2}{4}\sum\limits_{\lambda_q = \pm \frac{1}{2}} {\sum\limits_{\lambda_l = \pm \frac{1}{2}} {\left| \sum\limits_{n=0}^\infty  \tilde M_{\lambda_q \lambda_l}^{\left(n\right)} \right|^2 \left( {1 + 4\lambda_q \lambda_l \cos \theta^* } \right)^2 } }
\label{eq:sigmatotalamplitude_KK}
\end{equation}
where the complete amplitude consists of the SM term, exactly the term given in Eq~\ref{eq:totalamplitude}, plus an infinite KK tower of excitations with increasing mass,
\begin{equation}
\sum\limits_{n=0}^\infty \tilde M_{\lambda_q \lambda_l}^{\left(n\right)} \equiv \tilde M_{\lambda_q \lambda_l} + \sum\limits_{n=1}^\infty \tilde M_{\lambda_q \lambda_l}^{\left(n\right)}
\label{eq:onlytotalamplitude_KK}
\end{equation}
where $\tilde M_{\lambda_q \lambda_l}^{\left(0\right)}   \equiv   \tilde M_{\lambda_q \lambda_l}$ and where each contribution for $n>1$ can be written as 
\begin{equation}
	\tilde M_{\lambda_q \lambda_l}^{\left(n>0\right)}\left(\hat s \right) \equiv 
	\frac{e_q^{(n)} e_l^{(n)}}{\hat s - {\left(m_{\gamma^*}^{\left(n\right)}\right)}^2 + i \frac{\hat s}{m_{\gamma^*}^{\left(n\right)}} \sum\limits_{F}\Gamma_{\gamma^* \to F\bar F}^{\left(n\right)}} +
	\frac{g_{\lambda_q}^{\left(n\right)} g_{\lambda_l}^{\left(n\right)}}{\hat s - {\left(m_{Z^*}^{\left(n\right)}\right)}^2 + i \frac{\hat s}{m_{Z^*}^{\left(n\right)}} \sum\limits_{F}\Gamma_{Z^* \to F\bar F}^{\left(n\right)}}.
	\label{eq:totalamplitude_KK}
\end{equation}
Recalling Eq~\ref{eq:kk_mass}, the $n^{th}$ KK excitation masses $m_{Z^*}^{(n)}$ and $m_{\gamma^*}^{(n)}$ are given by
\begin{equation}
\begin{array}{rl}
	&m_{Z^*}^{(n)} = \sqrt{m_{Z^0}^2 + (n\cdot m^*)^2}\\
	&m_{\gamma^*}^{(n)} = n \cdot m^*.
\end{array}
\label{eq:KKmasses}
\end{equation}
As discussed previously, the current limits on the $m^*$ value are approaching 4 TeV and the LHC is expected to enable the expansion of the the search region.
Practically, the mass $m^*=$4 TeV is the value taken for the KK amplitudes as well as arbitrarily choosing an upper limit of $n=100$ which is large enough so that in practice, higher excitations do not play a significant r$\hat{\rm o}$le within the accessible LHC energy range.
Even though all the KK excitations higher than the first are beyond the reach of the LHC, their presence still affects the accessible LHC energy range due to interference contributions which are non-negligible even at energies far below their masses.
As mentioned, the couplings of the excited KK states to fermions are larger than the SM ones, Eq~\ref{eq:couplings}, by a factor of $\sqrt{2}$;
\begin{equation}
\begin{array}{rl}
	g_{\lambda_f}^{\left(n\right)} =
		\left\{
			\begin{array}{rl}
				g_{\lambda_f} &\mbox{if $n=0$} \\
				\sqrt{2} \cdot g_{\lambda_f} &\mbox{otherwise}
			\end{array}
		\right.\\\\
	e_f^{\left(n\right)} =
		\left\{
			\begin{array}{rl}
				e_f &\mbox{if $n=0$} \\
				\sqrt{2} \cdot e_f &\mbox{otherwise}
			\end{array}
		\right.
\end{array}
\label{eq:couplings_KK}
\end{equation}
where as in the previous section, the index $f$ can either represent the incoming quark or the outgoing lepton.
The sums $\sum\limits_F\Gamma_{{Z^*/\gamma^*} \to F\bar F}^{(n)}$ are the total  $\gamma^*$ and $Z^*$ decay widths to all fermion-antifermion pairs denoted by $F \bar F$.
Considering the SM terms from Equations~\ref{eq:width} and~\ref{eq:width_radiative_corrections}, these forms are affected by the $\sqrt{2}$ factor introduced in Eq~\ref{eq:couplings_KK}. 
In addition, with respect to the SM terms, a single power of the mass appearing in Eq~\ref{eq:width_radiative_corrections}, is replaced with the nominal mass (See Eq~\ref{eq:KKmasses}), starting from Eq~\ref{eq:width_radiative_corrections} for the $Z^*$ and from Eq~\ref{eq:width} for the $\gamma^*$,
\begin{equation}
\begin{array}{rl}
	\Gamma_{Z^* \to F\bar F}^{(n)} = \Gamma_{Z^0 \to F\bar F} \cdot
		\left\{
			\begin{array}{rl}
				1 											&\mbox{if $n=0$} \\
				2\frac{m_{Z^*}^{(n)}}{m_{Z^0}}	&\mbox{otherwise}
			\end{array} \right. \\\\
	\Gamma_{\gamma^* \to F\bar F}^{(n)} = \frac{N_C^F \alpha_{em} m_{\gamma^*}^{(n)}}{6} \cdot
		\left\{
			\begin{array}{rl}
				0 											&\mbox{if $n=0$} \\
				4e_F^2    &\mbox{otherwise}
			\end{array} \right.
\end{array}
\label{eq:KKwidths}
\end{equation}
This process can be realized as shown in Fig~\ref{fig:KK_feynman}.

Form the MC event generator stand point, in order to account for the entire KK tower -- namely a large enough $n$ -- the amplitude itself should be coded as a complex expression so that all the interference terms will emerge.
For instance, in the $\gamma/Z^0$ case in Pythia8, the amplitudes are pre-calculated and separated into the three different contributions ($\gamma$, $Z^0$ and interference).
Hence it is clear that in case of large number of diagrams, there will be many consequent interference terms and it is difficult to calculate and code.
\begin{figure}[!th]
\begin{center}
\begin{picture}(400,85)(0,0) 
\Text(0,20)[c]{$\sum\limits_{n=0}^\infty  \tilde M_{\lambda_q \lambda_l}^{\left(n\right)}=$}
\Text(40,40)[c]{$\bar q$}
\ArrowLine(55,20)(45,35) 
\Text(40,0)[c]{$q$}
\ArrowLine(45,5)(55,20) 
\Text(97,40)[c]{$l^+$}
\ArrowLine(90,35)(80,20) 
\Text(95,0)[c]{$l^-$}
\ArrowLine(80,20)(90,5) 
\Text(67.5,60)[c]{$m_\gamma^{\left(0\right)} = 0$}
\Photon(55,20)(80,20){3}{3.5}
\Text(67.5,10)[c]{$\gamma$}
\Text(45,20)[c]{$e_q$}
\Vertex(55,20){1.5}
\Text(92.5,20)[c]{$e_l$}
\Vertex(80,20){1.5}

\Text(110,80)[c]{$(n=0)$}
\Text(110,20)[c]{+}

\Text(125,40)[c]{$\bar q$}
\ArrowLine(140,20)(130,35) 
\Text(125,0)[c]{$q$}
\ArrowLine(130,5)(140,20) 
\Text(182,40)[c]{$l^+$}
\ArrowLine(175,35)(165,20) 
\Text(180,0)[c]{$l^-$}
\ArrowLine(165,20)(175,5) 
\Text(152.5,60)[c]{$m_{Z^0}^{\left(0\right)} = m_{Z^0}$}
\Photon(140,20)(165,20){3}{3.5}
\Text(152.5,10)[c]{$Z^0$}
\Text(129,20)[c]{$g_{\lambda_q}$}
\Vertex(140,20){1.5}
\Text(178,20)[c]{$g_{\lambda_l}$}
\Vertex(165,20){1.5}

\Text(193,20)[c]{+}

\Text(215,40)[c]{$\bar q$}
\ArrowLine(230,20)(220,35) 
\Text(215,0)[c]{$q$}
\ArrowLine(220,5)(230,20) 
\Text(272,40)[c]{$l^+$}
\ArrowLine(265,35)(255,20) 
\Text(270,0)[c]{$l^-$}
\ArrowLine(255,20)(265,5) 
\Text(242.5,60)[c]{$m_{\gamma^*}^{\left(1\right)} = m^*$}
\Photon(230,20)(255,20){3}{3.5}
\Text(242.5,10)[c]{$\gamma^*$}
\Text(212,20)[c]{$\sqrt{2}e_q$}
\Vertex(230,20){1.5}
\Text(271,20)[c]{$\sqrt{2}e_l$}
\Vertex(255,20){1.5}

\Text(291,80)[c]{$(n=1)$}
\Text(291,20)[c]{+}

\Text(320,40)[c]{$\bar q$}
\ArrowLine(335,20)(325,35) 
\Text(320,0)[c]{$q$}
\ArrowLine(325,5)(335,20) 
\Text(377,40)[c]{$l^+$}
\ArrowLine(370,35)(360,20) 
\Text(375,0)[c]{$l^-$}
\ArrowLine(360,20)(370,5) 
\Text(352.5,60)[c]{$m_{Z^*}^{\left(1\right)} = \sqrt{m_{Z^0}^2 + {m^*}^2}$}
\Photon(335,20)(360,20){3}{3.5}
\Text(352.5,10)[c]{$Z^*$}
\Text(315,20)[c]{$\sqrt{2}g_{\lambda_q}$}
\Vertex(335,20){1.5}
\Text(380,20)[c]{$\sqrt{2}g_{\lambda_l}$}
\Vertex(360,20){1.5}

\Text(405,20)[c]{+}
\Text(425,20)[c]{.\,.\,.}

\end{picture}
\caption{\textsl{The KK excitations tower of the gauge bosons $\gamma/Z^0$ starting from the $0^{th}$ SM state. Note the couplings and masses (which affect also the widths) of each level.}}
\label{fig:KK_feynman}
\end{center}
\end{figure}


\subsection{Analytic results}
In the following discussion, the general behavior of the cross section is studied under the LHC conditions, within the kinematic range of interest defined by several selection criteria.

Specifically, the di-muon final state, $pp \to \gamma^*/Z^* \to \mu^+\mu^-$, is chosen for this preliminary study.
The range of interest is determined by $p_T > 10$~GeV and $\left|\eta\right|<2.5$ which corresponds approximately to the $ATLAS$ trigger and detector acceptance.
To complete this definition, another threshold on the di-muon invariant mass is considered: $\sqrt{\hat s}>1$~TeV.

The restricted range implies that the hadronic distributions given in Equations~\ref{eq:sigmaLHC_total},~\ref{eq:sigmaLHCpdf},~\ref{eq:sigmaLHCpdf_y} and~\ref{eq:Afb_pdf} should be modified since they were obtained by integrating over $\cos\theta^*$ within $\left|\cos\theta^*\right| \leq 1$, whereas the effect of the cuts is restricting it to $\left|\cos\theta^*\right| \leq \left|\cos\theta_{\rm{max}}^*\right|$ given by
\begin{equation}
	\left|\cos\theta_{\rm{max}}^*\right| = \rm{min}\left\{ \left|\tanh\left( \eta^{\rm{cut}} - \left|y\right|\right)\right|\, ; \, \sqrt{1-\left(\frac{2 p^{\rm{cut}}_T}{\sqrt{\hat s}}\right)^2}  \right\}
\label{eq:costheta_max}
\end{equation}
where $\eta^{\rm{cut}}=2.5$, $p^{\rm{cut}}_T=10$~GeV, and $y$ is the rapidity of the lepton pair.
Due to these cuts, $y$ and $\sqrt{\hat s}$ are consequently confined to $|y|<2.5$ and $\sqrt{\hat s} \geq 2p_T^{\rm{cut}}$.
Using the same analytic approach, the two expressions in Eq~\ref{eq:costheta_max} can be inverted to obtain the effective cuts on $\hat s$.

\begin{figure}[!th]
\centering
\includegraphics[scale=0.85,angle=0]{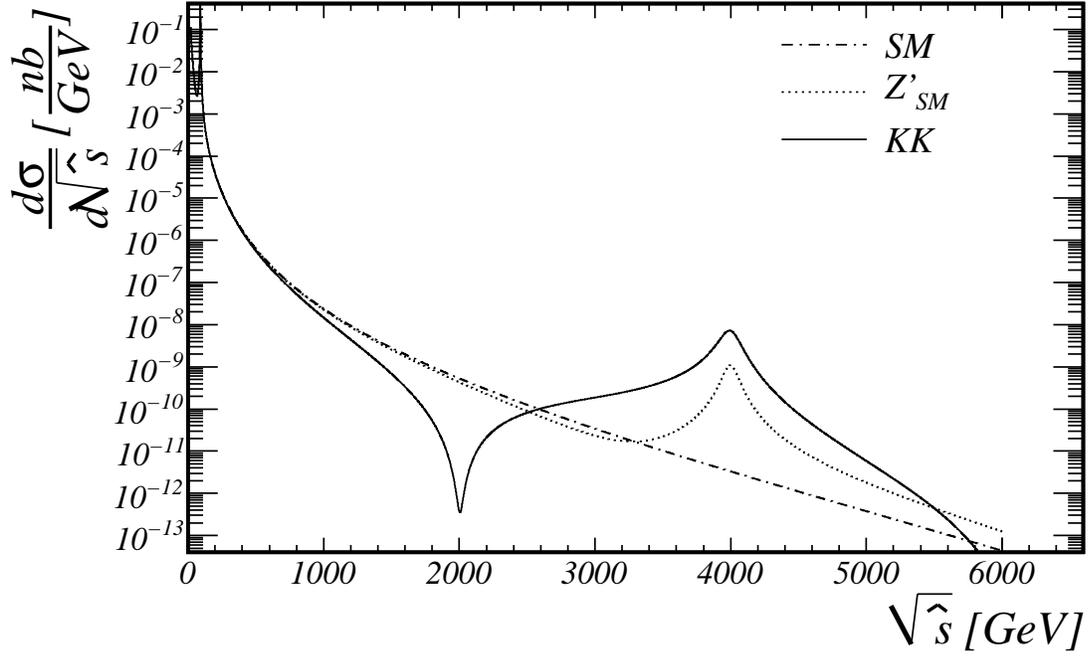}
\caption{\textsl{The invariant-mass distributions of the three models, KK (solid), $Z'_{{\rm{SM}}}$ (dotted) and SM (dash-dot) within the kinematic region described in the text.}}
\label{fig:invMass_theory_3models}
\end{figure}

The $\sqrt{\hat s}$ distributions around the first excitation are shown in Fig~\ref{fig:invMass_theory_3models} where a strong destructive interference is clearly seen between 1 and 2~TeV for the KK line.
The distributions of the SM and the $Z'_{\rm{SM}}$ model discussed in the previous section are also shown for comparison.
As mentioned in the introduction, the shapes from Fig~\ref{fig:invMass_theory_3models} could enable the discrimination between the models already at $\mathcal{L}=100$ fb$^{-1}$.

An estimation for the number of events that are expected to be measured by the $ATLAS$ detector in the range of interest is necessary for generating the MC pseudo-data samples with a realistic size.
The mass range considered is, $1 \leq \sqrt{\hat s} \leq 6$~TeV with integrated luminosity of $\mathcal{L}=$100~fb$^{-1}$.
After applying the selection cuts a total number of $\sim$400 KK events are expected in that overall mass range at the respective luminosity and $\sim$190 KK events in the KK peak area, $2 \leq \sqrt{\hat s} \leq 5$~TeV, see Table~\ref{table:expectedEvents}.
The number of events in the overall mass range of interest, $1 \leq \sqrt{\hat s} \leq 6$~TeV, will be used as an input for the simulation.

\begin{figure}[!th]
\centering
\includegraphics[scale=0.85,angle=0]{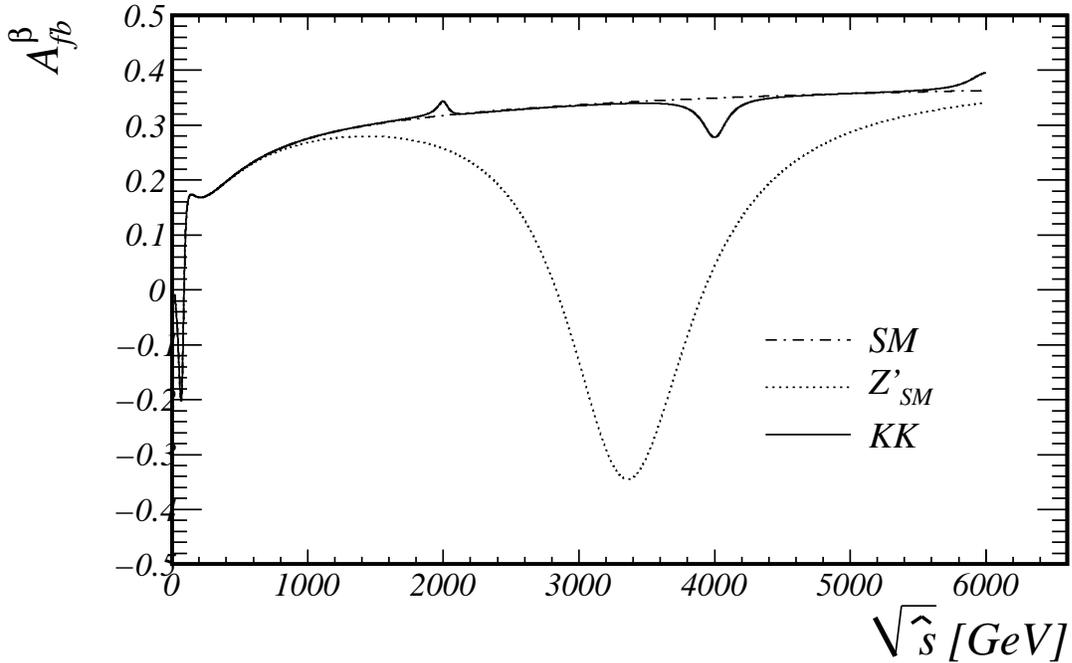}
\caption{\textsl{The forward-backward asymmetry within the kinematic region described in the text, corresponding to the three models, KK (solid), $Z'_{{\rm{SM}}}$ (dotted) and SM (dash-dot).}}
\label{fig:theoreticalasymmetry}
\end{figure}

The analytical function for the  forward-backward asymmetry relative to the boost direction, denoted by $A_{fb}^{\beta}$, is shown in Fig~\ref{fig:theoreticalasymmetry}. 
Large differences between the forward-backward asymmetry for the $Z'_{{\rm{SM}}}$ and the KK models around the KK resonance are apparent.
Depending on the integrated luminosity, these expected differences, together with the observation of a significant peak, might enable the discrimination between the three models.
As illustrated in Table~\ref{table:expectedEvents}, the statistics in this range of masses is expected to be poor and as such, calculating the asymmetry in a wider region might be necessary.
To give a prediction for $A_{fb}^{\beta}$ in $2 \leq \sqrt{\hat s} \leq 5$~TeV, the theoretical function is averaged over $\sqrt{\hat s}$,
\begin{equation}
	\widehat{A_{fb}^{\beta}} \equiv \int\frac{d\sigma}{d\hat s} A_{fb}^{\beta} d\hat s \times \left(\int\frac{d\sigma}{d\hat s}d\hat s\right)^{-1}.
\label{eq:averagedAsymmetry}
\end{equation}
The predicted values for this averaged asymmetry, $\widehat{A_{fb}^{\beta}}$, in LO are summarized in Table~\ref{table:expectedAsymmetry}.
\vspace{-4mm}
\begin{table}[!th]
\centering
\caption{\textsl{The expected number of events in $\mathcal{L}$=100 fb$^{-1}$ for the three models, SM, $Z'_{{\rm{SM}}}$ and KK for $1 \leq \sqrt{\hat s} \leq 6$~TeV and for $2 \leq \sqrt{\hat s} \leq 5$~TeV within the described kinematic region.}}
\vspace{4mm}
\begin{tabular}{ccc}\hline\hline
	Model	& \hspace{10mm} Events in $1 \leq \sqrt{\hat s}\leq 6$ TeV & \hspace{10mm} Events in $2 \leq \sqrt{\hat s} \leq 5$ TeV	\\
	\hline
	SM						& \hspace{10mm} $\sim 480$		&	\hspace{10mm} $\sim 15$		\\
	$Z'_{{\rm{SM}}}$	& \hspace{10mm} $\sim 460$		&	\hspace{10mm} $\sim 30$		\\
	KK						& \hspace{10mm} $\sim 400$		&	\hspace{10mm} $\sim 190$	\\
	\hline\hline
	\end{tabular}
	\label{table:expectedEvents}
\end{table}
\vspace{-4mm}
\begin{table}[!th]
\centering
\caption{\textsl{The expected values of $\widehat{A_{fb}^{\beta}}$ in LO for the three models SM, $Z'_{{\rm{SM}}}$ and KK around the KK resonance $2 \leq \sqrt{\hat s} \leq 5$ TeV within the described kinematic region.}}
\vspace{4mm}
\begin{tabular}{cc}\hline\hline
	Model & \hspace{15mm} $\widehat{A_{fb}^{\beta}}$ in $2 \leq \sqrt{\hat s} \leq 5$ TeV\\
	\hline
	SM						& \hspace{15mm} $0.325$\\
	$Z'_{{\rm{SM}}}$	& \hspace{15mm} $0.090$\\
	KK						& \hspace{15mm} $0.308$\\
	\hline\hline
	\end{tabular}
	\label{table:expectedAsymmetry}
\end{table}


\clearpage

\subsection{Strategy of the experimental kinematics study}
This sub-section presents some preliminary results obtained using \mosesWithSpace for the KK process with the di-muon final state, $pp \to \gamma^*/Z^* \to \mu^+\mu^- + X$, at the generator level.
The results from the KK process are presented and compared with the SM and the $Z'_{{\rm{SM}}}$ results (with the same final state) with an emphasis on the discrimination between the possible KK or $Z'_{{\rm{SM}}}$ signals.

The realization of the processes within the LHC and $ATLAS$ kinematic regions enables observables to be studied in a realistic regime at the generator level.
The next stage is to embed the new subprocesses within fully simulated events including the effects of initial and final state radiation (ISR and FSR), parton showering, hadronization, proton remnant fragmentation, particle decay etc. All these effects are available within Pythia8.
Because of factorization, only the hard subprocess needs to be generated, and integrated within Pythia8.
Therefore, Pythia8 itself is responsible for the subsequent forward evolution of the event from the time and energy scale of the provided hard-process.
It is also responsible for the backwards evolution to the initial conditions of the interacting protons.\\

The analysis starts from the measured di-lepton final state, namely, the 4-momenta of the charged leptons $p_{l}^{\nu}$ and $p_{\bar l}^{\nu}$ given in the colliding protons CM frame, denoted by $\mathcal{O}$.
The sum  of these 4-momenta measured in the $\mathcal{O}$ frame is
\begin{equation}
	Q^{\nu} =p_{l}^{\nu} + p_{\bar l}^{\nu}\\
\label{eq:Q}
\end{equation}
It is useful to define the di-lepton squared invariant mass, $Q^2 = Q_{\nu}Q^{\nu} = \hat s$ and its transverse momentum, $\vec{Q_T} = \left(p_{l}^1+p_{\bar l}^1, p_{l}^2+p_{\bar l}^2\right)$.
The complementary components, $Q^{\pm} = \frac{1}{\sqrt{2}}\left(Q^0 \pm Q^3\right)$, can also be defined
It is useful to write the rapidity of the di-lepton system $y_Q = \frac{1}{2}\ln\left(\frac{Q^+}{Q^-}\right)$.
For the following discussion the masses of the leptons and partons are small with respect to $Q = \sqrt{\hat s}$ and are thus neglected, as are the proton masses which are small with respect to $\sqrt{s}$.

\begin{figure}[!th]
\centering
	\begin{overpic}[scale=0.81]{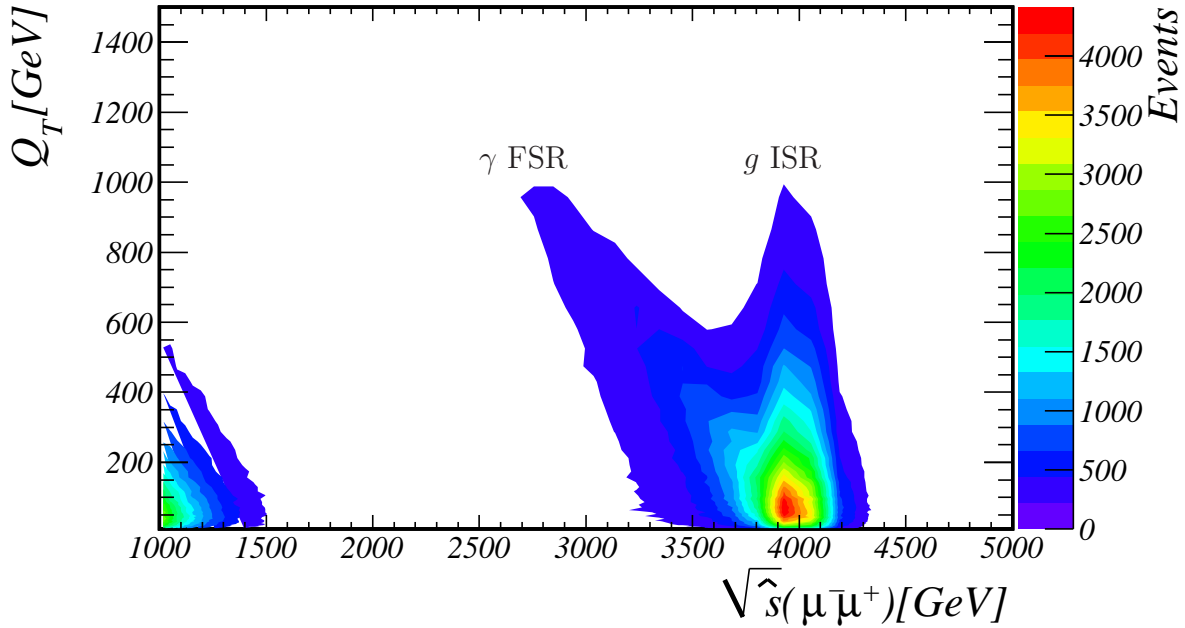}
		\put(290,175){$g$ ISR}
		\put(190,175){$\gamma$ FSR}
	\end{overpic}
\caption{\textsl{The $Q_T$ vs. $\sqrt{\hat s}$ distribution of the exchanged KK bosons. The separate contributions of ISR and FSR are clearly seen (See text).}}
\label{fig:QTvsMhat_with_ISRandFSR}
\end{figure}

In the LO Drell-Yan picture, both original partons have momentum only along the original $z$ direction of the $\mathcal{O}$ frame.
Correspondingly, the di-lepton system CM can have momentum only along the same direction.
When considering higher orders in the Drell-Yan picture, if one of the original partons emits an ISR gluon it can obtain momentum also in the transverse direction.
The intermediate state and the di-lepton system will then have a momentum component in the transverse direction.
In general this will modify the angular distribution of the outgoing leptons with respect to that discussed earlier.
In addition, it is also possible for one of the outgoing leptons to radiate an FSR photon. 
If it is not detected and corrected for in the lepton kinematics, then the resulting invariant mass of the di-lepton system will be shifted towards lower values.
The effects of the ISR and the FSR can be demonstrated in Fig~\ref{fig:QTvsMhat_with_ISRandFSR}.

The unique 2$d$ distribution seen in Fig~\ref{fig:QTvsMhat_with_ISRandFSR} is the first comprehensive result of the entire \mosesWithSpace and Pythia8 framework that is being presented.
It illustrates several aspects of the KK process which can be seen in it very clearly.
To obtain this plot, a sample of 10$^6$ KK full events was generated using \mosesWithSpace and Pythia8, including all steps of the complete event generation.
The events were selected such that they are within the kinematic range of interest introduced in the previous sub-section.
The KK resonance can be seen around $\sqrt{\hat s}\sim$4~TeV with two distinct $Q_T$ tails, one vertical which is associated with the ISR effect and the other one diagonal which is associated with the FSR effect.
This picture, of initial and final state radiation, piles up more complication as can be clarified in the next discussion.

Thus, from the $Q_T$ distribution of the di-lepton system (with respect to $\mathcal{O}$ frame), it can be clearly seen that the di-lepton CM has a non-negligible momentum in the transverse direction due to the radiation of the initial state gluon or final state photon.
The tails seen in the $Q_T$ distribution in Fig~\ref{fig:QTvsMhat_with_ISRandFSR} can be rather large, causing once more a problem for the way $\cos\theta^*$ is defined.

Following the results seen in Fig~\ref{fig:QTvsMhat_with_ISRandFSR}, if the FSR switch in Pythia8 is turned off then the related $Q_T$ diagonal tail vanishes, as expected, and no corresponding shift in the invariant mass occurs.
In addition, turning off the ISR as well, then the vertical $Q_T$ tail does not extend significantly above at $Q_T \sim$10~GeV rather than the $\sim$1~TeV shown in Fig~\ref{fig:QTvsMhat_with_ISRandFSR}.
However, further consideration of these phenomena suggests that they may not have a significant effect on the overall kinematic distributions.

For di-muon events where the FSR photon has sufficiently high transverse energy, the measured muon might fail the $p_T > 10$~GeV cut and this would affect the distribution.
However, in this case, the photon itself should be sufficiently well separated from the muon so that it can be detected independently.
On the other hand, for di-electron events with a photon being sufficiently collinear with the electron, the calorimeter cluster would include both the electron and photon and the reconstructed energy would correspond well to the initial electron energy before radiation.
In any real study, the effect of final state photon radiation can be simulated and the effect on the trigger and selection efficiency can be investigated.
However, this case will not be discussed further here.
The case of ISR gluon requires more attention and will be discussed in the next section.

\subsubsection{Handling with the ISR effect}
To consistently handle the effects of ISR, since the boost direction no longer coincides with the beam axis, it is not clear which axis should be used for the measurement of $\cos\theta^*$.
There are two approaches to deal with that problem;
\begin{itemize}
\item minimize the ISR and FSR effects by applying a cut on large $Q_T$'s ({\em ie}, $Q_T < Q_T^{\rm{max}} \sim \mathit{O}$(100~GeV)), 
where for most of these events, the di-muon system momentum will be along the beam axis.
In this case, the LO formalism where there is no ISR effect is valid and can be used.
However this will reduce the number of events in the sample.
\item analyze the events while taking into account the ISR, by rotating the di-lepton rest frame by some angle so that the $z$-axis will 
be slightly modified.
A common choice is the Collins-Soper (CS) reference frame~\cite{COLLINSSOPER,COLLINSSOPERZ0,COLLINSSOPERZ0W}, denoted here by $\mathcal{O}'$.
It can be shown that this choice, minimizes the contribution from longitudinally polarized $\gamma^*/Z^*$ which can be produced due to the gluon ISR and thus, can affect the angular distribution.
\end{itemize}
Note that in the $\mathcal{O}'$ frame, $\cos\theta '$ again has the sign ambiguity previously discussed.
This ambiguity results from our arbitrary selection of the original $z$ direction in the $\mathcal{O}$ frame.
This implies that a reclassification of $\cos\theta '$ can be performed with respect to the rapidity sign such that if $y_Q < 0$ then $\cos\theta ' \to -\cos\theta '$.
This is possible since in most of the events, the CS rotation angle will be small enough so the choice of the $z'$ axis will point approximately in the direction of one of the protons ($\vec{p'}_{\rm{p}_1}$ or $\vec{p'}_{\rm{p}_2}$).

The lepton $\cos\theta '$ distribution to first order in $\mathcal{O}(\alpha_s)$ must have the form~\cite{COLLINSSOPERZ0}
\begin{equation}
	\frac{1}{N}\frac{dN}{d\cos\theta '} = \frac{3}{8}\left[1 + \frac{1}{2}A_0 + A_4\cos\theta ' + \left(1-\frac{3}{2}A_0\right)\cos^2\theta ' \right]
\label{eq:costhetadistribution_new}
\end{equation}
where the coefficients $A_0$ and $A_4$ can be functions of the kinematic variables $s$, $\hat s$, $y_Q$ and $Q_T$, and where for each event $\cos\theta '$ should be calculated relative to the sign of $y_Q$.
Finally, the forward-backward asymmetry in the CS $\mathcal{O}'$ frame can be realized from the last expression as $A_{fb}^{{\rm{CS}}} = \frac{3}{8}A_4$ and, as expected, now depends on both  $\hat s$ and $Q_T$.
The resulting forward-backward asymmetry should be unaffected by the gluon ISR and the consequent transformation to the CS frame, so the only affected part is the symmetric one that involves $A_0$.
This coefficient is relevant for the examination of the complete $\cos\theta '$ distribution.

The angular distribution given in Eq~\ref{eq:costhetadistribution_new} corresponds to an exchanged particle with spin-1 and in that context, the $\cos\theta '$ distribution of a spin-2 particle such as the graviton, exhibits a completely different behavior~\cite{GRAVITON,GRAVITON3,GRAVITON2}.
As mentioned in the introduction, a spin classification should be also done.

\clearpage

\subsubsection{The motivation for extracting the angular coefficients}
The motivation for extracting the $A_4$ coefficient is the insight it can provide on the couplings of the new heavy gauge bosons to the SM fermion fields.
Apart from that, there is a possibility that this coefficient may support the discrimination between the two spin-1 models, as well as a spin-2 model.
However, it is clear that extracting the coefficients from the pseudo-data and comparing with the MC reference coefficients, can give no more discrimination than comparing the distributions themselves.

As illustrated in Table~\ref{table:expectedAsymmetry} and in Fig~\ref{fig:theoreticalasymmetry}, the two non-SM forward-backward asymmetry coefficients differ significantly for any mass in the range $2 \leq \sqrt{\hat s} \leq 5$ TeV.
However, the values given in Table~\ref{table:expectedAsymmetry}, considered as an analytic results, are averaged and also not strictly accurate for the following analysis since the ISR (and FSR) effect was not included.
In the case of ISR, the measured asymmetry is expected to slightly change and in addition, there is another coefficient, the symmetric $A_0$ (vanishing in the case of no ISR).
Thus, the best estimation for these coefficients, under the influence of ISR, can come from a high statistics MC reference sample\footnote{high statistics compared to the expected between 1~TeV and 6~TeV, see Table~\ref{table:expectedEvents}} (for each model).
The coefficients can be extracted by fitting the $\cos\theta '$ distribution (given in Eq~\ref{eq:costhetadistribution_new}) corresponding to the high statistics MC reference sample and this is done also for the realistic statistics (e.g. pseudo-data with $\mathcal{L} = 100$~fb$^{-1}$) to quantify how well it might work.
In that sense, it should be stressed that although the $A_0$ coefficient is relevant for the complete $\cos\theta '$ distribution study, its role here is no more than to technically improve the fit whose primary objective is the forward-backward asymmetry coefficient.
The cuts on $p_T$, $\eta$ and $\sqrt{\hat s}$, introduced in the previous section, are applied on all the samples.

\subsubsection{The discrimination between the spin-1 models}
As mentioned in the introduction section, having the large MC reference samples and the $\mathcal{L}$=100(500) fb$^{-1}$ pseudo-data samples, the comparison can be performed using the Kolmogorov test for the $\sqrt{\hat s}$ and the $\cos\theta '$ distributions.
Each distribution exhibits a special characteristic behavior for the examined model.
In that context, the Kolmogorov test can also provide good sensitivity for the determination of the spin of the exchanged particle.

\subsubsection{The strategy}
The strategy used to classify the observed signal, can be summarized in four points:
\begin{enumerate}
\item searching for a significant resonance in the di-muon $\sqrt{\hat s}$ distribution above the small expected SM background.
\item comparing the $\cos\theta '$ distribution around the resonance to a spin-1 and spin-2 resonance distributions.
This step is not shown in this paper\footnote{Optionally, the azimuthal angle distributions can also be used for this purpose\cite{AZIMUTHALRS}.}.
\end{enumerate}
assuming a spin-1 resonance,
\begin{enumerate}
\item[3.] comparing the $\sqrt{\hat s}$ and $\cos\theta '$ distributions to all three MC reference distributions.
\item[4.] fitting the $\cos\theta '$ distribution from Eq~\ref{eq:costhetadistribution_new} to the data.
\end{enumerate}
In the next sub-section, a collection of results obtained from the preliminary analysis at the generator level are gathered and shown in detail.


\subsection{A collection of the results}
In the first part of this subsection, a detailed discussion is given on the Maximum Likelihood (ML) fit results for the $\cos\theta '$ distributions.
In the second part, the Kolmogorov test results for the $\cos\theta '$ and $\sqrt{\hat s}$ distributions are summarized with detailed comments on the possibilities of discriminations.

\begin{figure}[!th]
	\centering
	\begin{overpic}[scale=0.85]{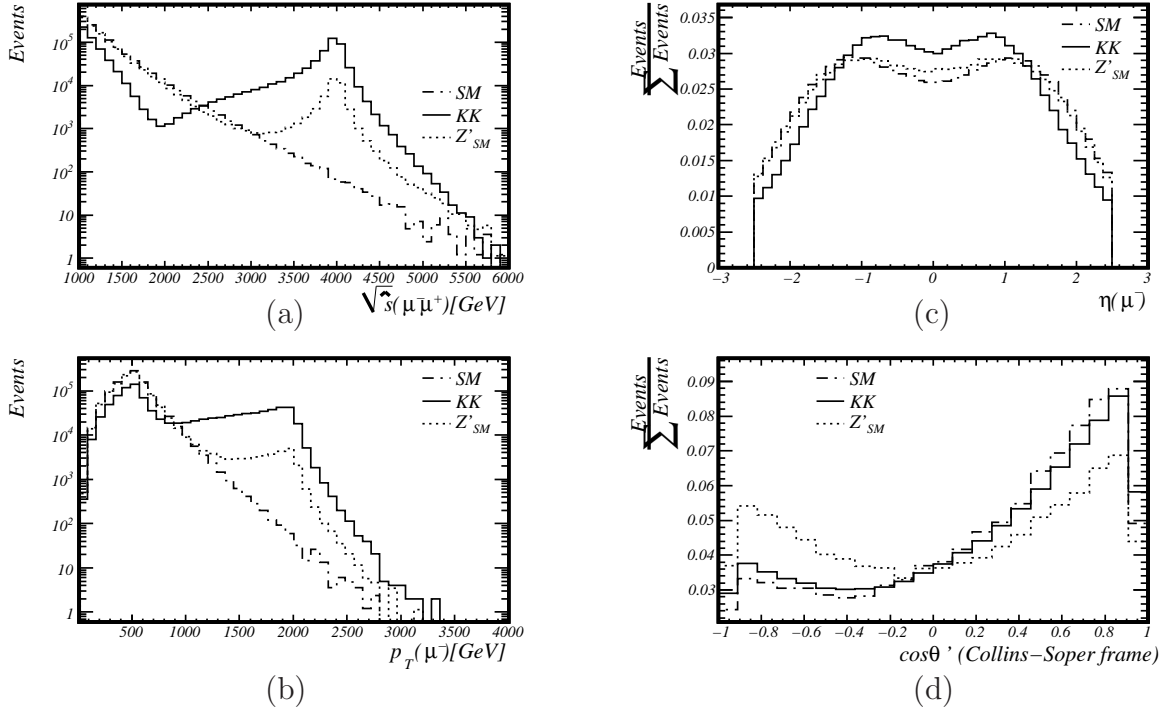}
		\put(110,135){(a)}
		\put(110,-7){(b)}
		\put(355,135){(c)}
		\put(355,-7){(d)}
	\end{overpic}
\caption{\textsl{Kinematic distributions of the MC reference samples for the KK (solid), the SM (dash-dot) and the $Z'_{{\rm{SM}}}$ (dotted) models. {\em (a)} The di-muon $\sqrt{\hat s}$ distribution, {\em (b)} the muons $p_T$ distribution, {\em (c)} the muons normalized $\eta$ distribution and, {\em (d)} the normalized muons $\cos\theta '$ distribution.}}
\label{fig:KK_kinematics_highLuminosity}
\end{figure}

\begin{figure}[!th]
\centering
	\begin{overpic}[scale=0.85]{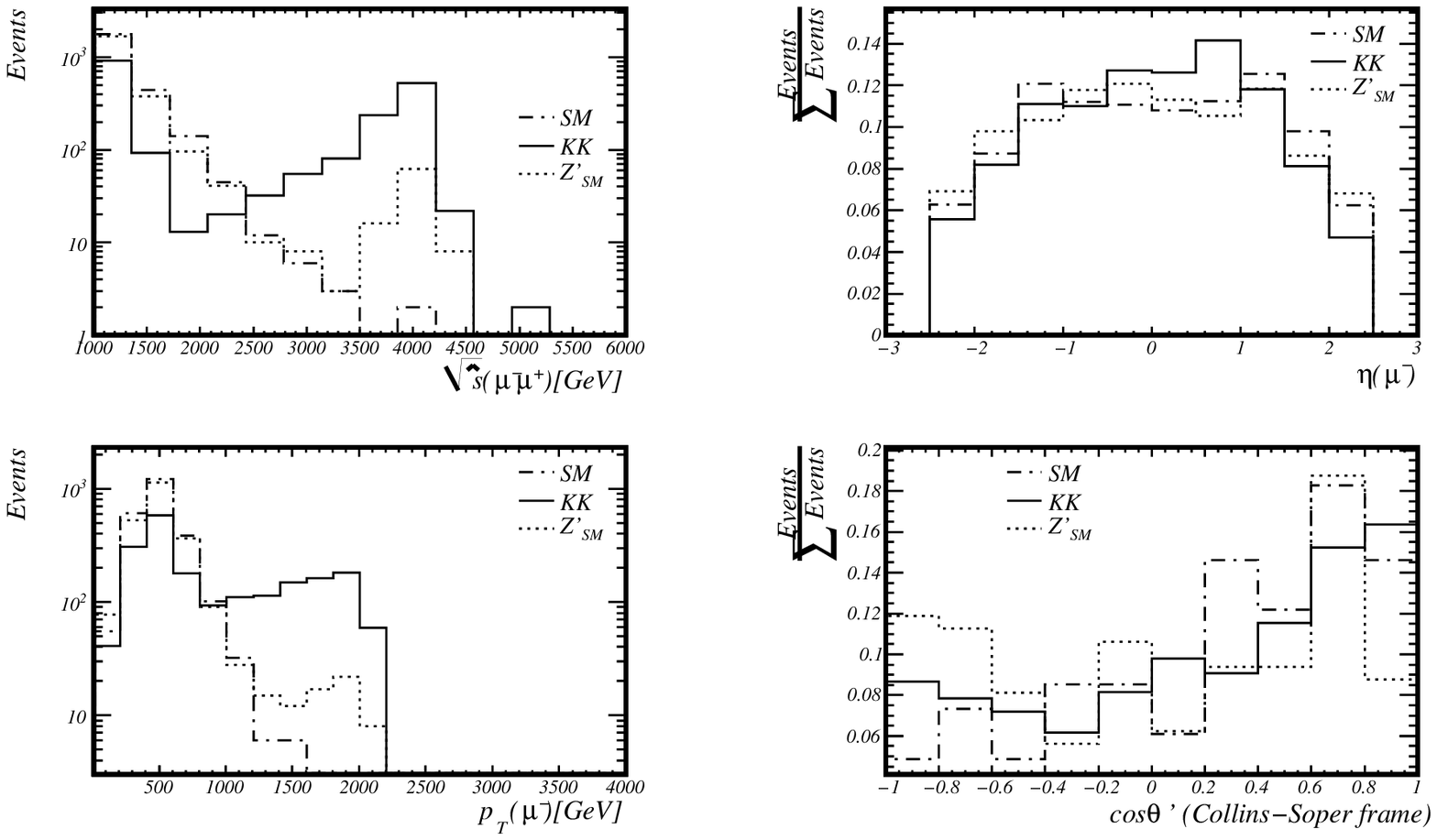}
		\put(110,135){(a)}
		\put(110,-7){(b)}
		\put(355,135){(c)}
		\put(355,-7){(d)}
	\end{overpic}
\caption{\textsl{Kinematic distributions of the $\mathcal{L}$=500 fb$^{-1}$ samples. See Fig~\ref{fig:KK_kinematics_highLuminosity}}}
\label{fig:KK_kinematics_lowLuminosity500}
\vspace{10mm}
	\begin{overpic}[scale=0.85]{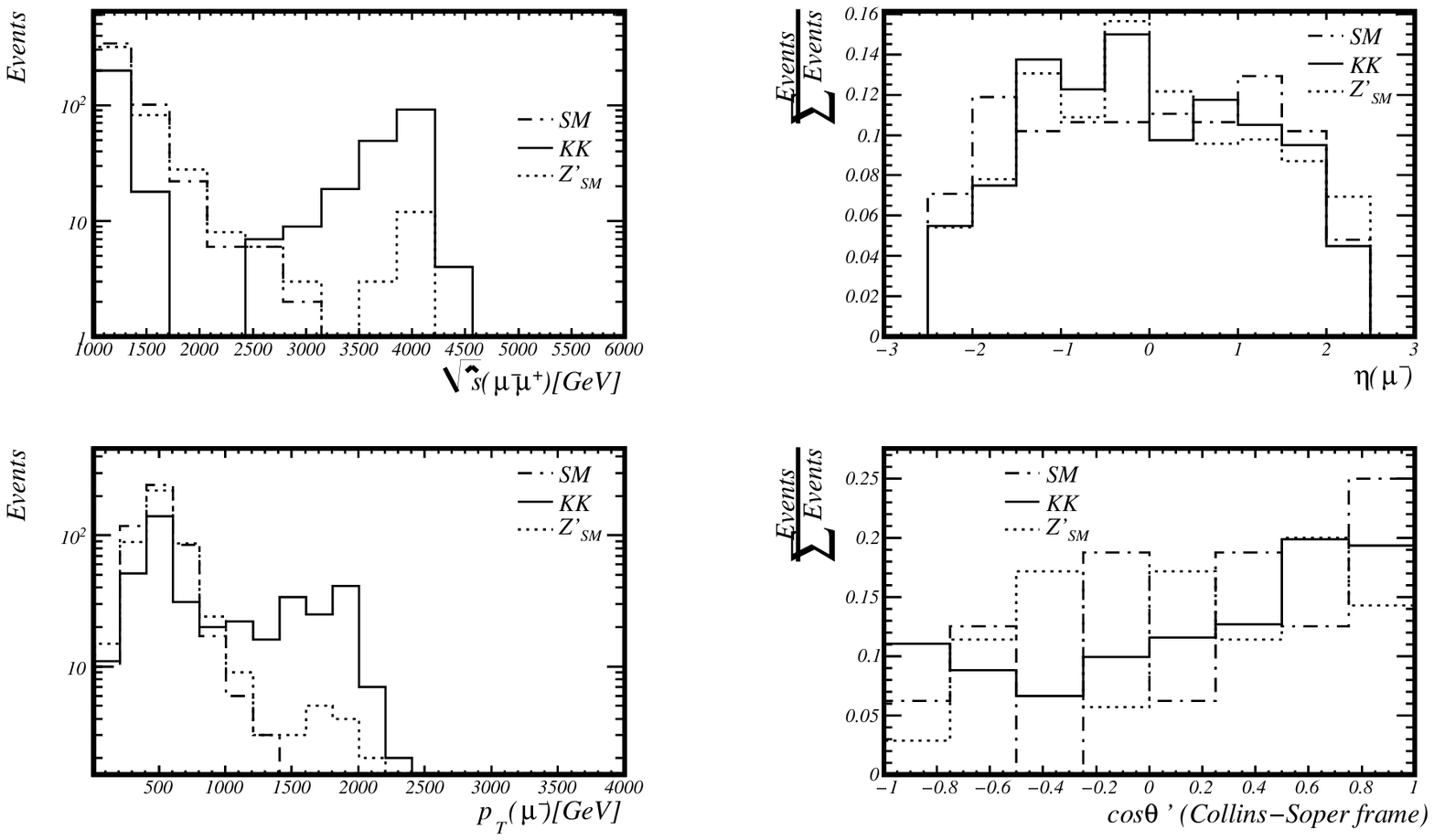}
		\put(110,135){(a)}
		\put(110,-7){(b)}
		\put(355,135){(c)}
		\put(355,-7){(d)}
	\end{overpic}
\caption{\textsl{Kinematic distributions of the $\mathcal{L}$=100 fb$^{-1}$ samples. See Fig~\ref{fig:KK_kinematics_highLuminosity}}}
\label{fig:KK_kinematics_lowLuminosity100}
\end{figure}

First, let the $\sqrt{\hat s}$, $p_T$, $\eta$ and $\cos\theta '$ distributions, for the three models, be gathered and shown at high and low statistics. 
The kinematic distributions of the large MC reference samples are shown in Fig~\ref{fig:KK_kinematics_highLuminosity} where the distributions of the pseudo-data samples can be seen in Fig~\ref{fig:KK_kinematics_lowLuminosity500} and~\ref{fig:KK_kinematics_lowLuminosity100} respectively.

In Fig~\ref{fig:KK_kinematics_highLuminosity}, the sizes (luminosities) of the SM and the $Z'_{{\rm{SM}}}$ MC reference samples are normalized to the arbitrarily chosen size (luminosity) of the KK reference sample ($10^6$ KK events).
In Figures~\ref{fig:KK_kinematics_lowLuminosity500} and~\ref{fig:KK_kinematics_lowLuminosity100} the sizes of all three samples are determined by the two LHC luminosities, {\em ie} $\mathcal{L}$=500 fb$^{-1}$ and $\mathcal{L}$=100 fb$^{-1}$.
These three figures are arranged in the same format so their properties are identical.

\clearpage

\subsubsection[The ML fit for the spin-1 $\cos\theta '$ distribution]{The ML fit for the spin-1 \boldmath$\cos\theta '$ distribution}
\begin{figure}[!th]
\centering
\includegraphics[scale=0.85,angle=0]{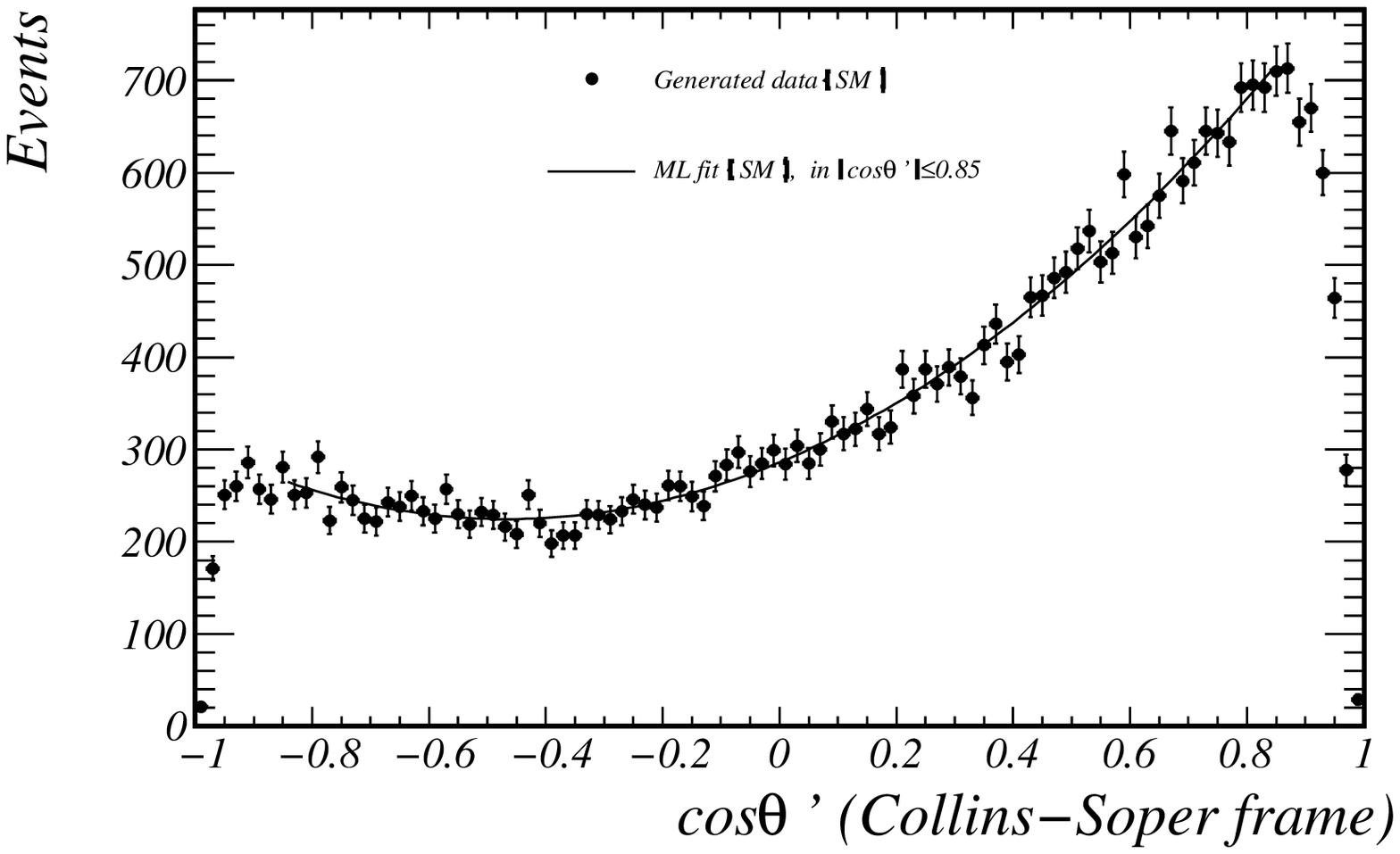}
\caption{\textsl{The SM $\cos\theta '$ distribution in the range $2 \leq \sqrt{\hat s} \leq 5$ TeV vs. the ML fit.}}
\label{fig:SM_fit}
\end{figure}

\begin{figure}[!th]
\centering
\includegraphics[scale=0.85,angle=0]{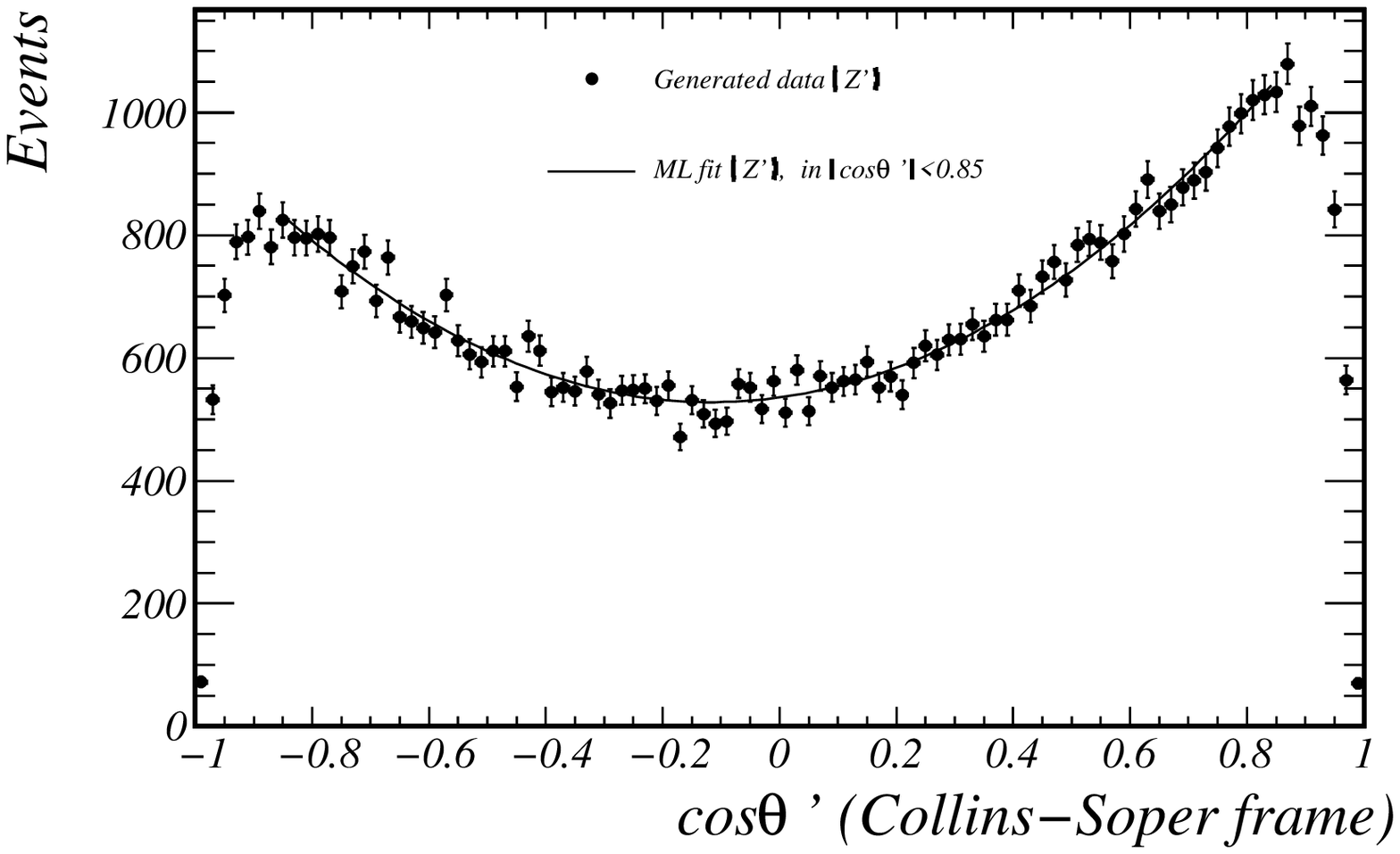}
\caption{\textsl{The $Z'_{{\rm{SM}}}$ $\cos\theta '$ distribution in the range $2 \leq \sqrt{\hat s} \leq 5$ TeV vs. the ML fit.}}
\label{fig:E6_fit}
\end{figure}

\begin{figure}[!th]
\centering
\includegraphics[scale=0.85,angle=0]{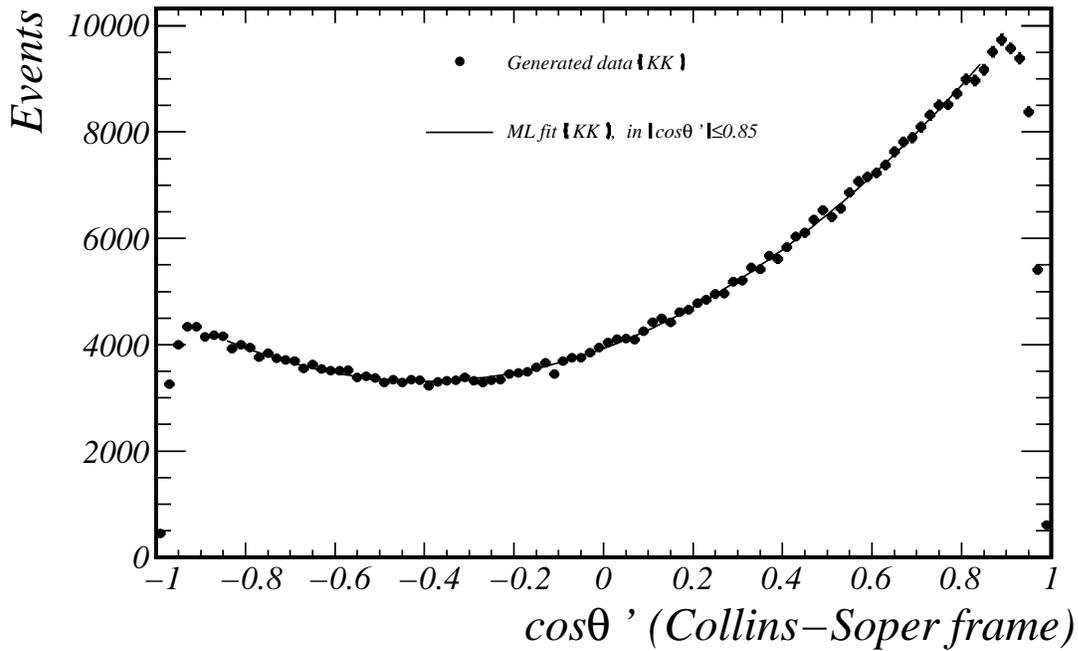}
\caption{\textsl{The KK $\cos\theta '$ distribution in the range $2 \leq \sqrt{\hat s} \leq 5$ TeV vs. the ML fit.}}
\label{fig:KK_fit}
\end{figure}

The ML fit procedure, on one hand, should be unbiased by the influence of the specific kinematic range of interest and by the choice of a specific model.
On the other hand, due for poor statistics one will want to use all events that are in the kinematic region.
In other words, one is left with two possibilities:
\begin{enumerate}
\item performing the fit in the entire allowed range of $\cos\theta '$ in order not to lose events at the chopped edges of the distribution\footnote{see the chopped edges of the $\cos\theta '$ distribution in Fig~\ref{fig:SM_fit},~\ref{fig:E6_fit} and~\ref{fig:KK_fit}}.
The described kinematic region along with the choice of a certain model out of the three, will give a biased fit result.
This bias will appear due to the fact that the angular acceptance, (which should be explicitly modeled in the fit), depends on the $A_{fb}^{{\rm{CS}}}$ and thus, it depends on the choice of the model.
Therefore, a detailed study of the dependence of the fit result for $A_{fb}^{{\rm{CS}}}$ on its expected value should be performed.
This study should be done for various values of $A_{fb}^{{\rm{CS}}}$ so eventually one can understand this dependence and correct the fit result.
\item performing the fit in a smaller range of $\cos\theta '$ where there are no edge effects due to the limited kinematic range of interest.
This way part of the event sample is lost but there is no need to introduce a prior correction to compensate the bias tendency.
\end{enumerate}
Using the event-by-event ML method, the second option is chosen to fit the CS distribution (from Eq~\ref{eq:costhetadistribution_new}) to the large MC reference sample and the two pseudo-data samples.
The range of the fit is limited to $\left|\cos\theta '\right| \leq 0.85$ and that enables to perform the fit without any bias.
The ML fit~\cite{ROOT} results for the $A_0$ and $A_{fb}^{{\rm{CS}}}$ coefficients of the MC reference samples and the pseudo-data ($\mathcal{L}$=100(500) fb$^{-1}$) are summarized in Table~\ref{table:angularFit}.
The fitted curves of the MC reference samples are shown in Figures~\ref{fig:SM_fit},~\ref{fig:E6_fit} and~\ref{fig:KK_fit}.
\begin{table}[!th]
\centering
\caption{\textsl{A summary of the ML fit results for the $\cos\theta '$ distributions in the CS $\mathcal{O}'$ frame around the KK resonance $2 \leq \sqrt{\hat s} \leq 5$ TeV and within the interval $\left|\cos\theta '\right| \leq 0.85$.}}
\vspace{4mm}
\begin{tabular}{ccccc}
\hline\hline
\multirow{2}{*}{}
	& \,			& MC 				& $\mathcal{L}$=500 fb$^{-1}$ 	& $\mathcal{L}$=100 fb$^{-1}$\\
	& Model	 	& reference		& pseudo-data 		& pseudo-data\\
\hline\hline
\multirow{3}{*}{${A_{fb}}$}
	& SM 						& $0.3476 \pm 0.0053$ 	& $0.37 \pm 0.11$			& $0.48 \pm 0.25$ \\
	& $Z'_{\rm{SM}}$ 		& $0.0900 \pm 0.0042$ 	& $0.117 \pm 0.083$		& $0.24 \pm 0.18$ \\
	& KK 						& $0.2958 \pm 0.0015$ 	& $0.216 \pm 0.035$		& $0.304 \pm 0.078$ \\
\hline
\multirow{3}{*}{${A_0}$}
	& SM 						& $0.00 \pm 0.02$ 		& $0.09 \pm 0.39$ 	& $0.62 \pm 1.40$ \\
	& $Z'_{\rm{SM}}$		& $-0.025 \pm 0.015$ 	& $-0.53 \pm 0.26$ 	& $1.06 \pm 0.86$ \\
	& KK 				 		& $0.0075 \pm 0.0056$ 	& $0.00 \pm 0.13$ 	& $0.00 \pm 0.28$ \\
\hline
\multirow{3}{*}{Events}
	& SM 						& 30145 	& 73 	& 13 \\
	& $Z'_{\rm{SM}}$ 		& 57053 	& 142 & 31 \\
	& KK 						& 413391	& 798 & 150 \\
\hline
\hline
\end{tabular}
\label{table:angularFit}
\end{table}

\clearpage

\subsubsection[The Kolmogorov test for the $\cos\theta '$ and $\sqrt{\hat s}$ distributions]{The Kolmogorov test for the \boldmath$\cos\theta '$ and $\boldsymbol{\sqrt{\hat s}}$ distributions}
The returned value of the Kolmogorov test is its probability, {\em ie}, a value much less than one means NOT compatible~\cite{ROOT}.
The unbinned Kolmogorov test is chosen for two reasons; first, at low statistics it is usually better to perform an unbinned analysis and second, it does not add more arbitrary systematics that have to be studied due to sensitivity to arbitrary binning choices.
The unbinned test results are summarized in Tables~\ref{table:angularKolmogorov} and~\ref{table:mHatKolmogorov}.
\begin{table}[!th]
\centering
\caption{\textsl{A summary of the unbinned Kolmogorov test results for the $\cos\theta '$ distributions in the CS $\mathcal{O}'$ frame within the interval $2 \leq \sqrt{\hat s} \leq 5$ TeV and in the full $\cos\theta '$ interval ([-1,1]).}}
\vspace{4mm}
\begin{tabular}{ccc||ccc}
\hline\hline
	MC ref'  	& Luminosity		& \#Events of		& SM				& $Z'_{\rm{SM}}$ 	& KK\\
	model 			& of pseudo-data 	& pseudo-data	& pseudo-data	& pseudo-data	& pseudo-data\\
\hline\hline
\multirow{2}{*}{SM}
	& $\mathcal{L}$=500 fb$^{-1}$ & 82 & 0.97  & 0.002 & 0.008 \\
	& $\mathcal{L}$=100 fb$^{-1}$ & 16 & 0.887 & 0.66  & 0.89 \\
\hline
\multirow{2}{*}{$Z'_{\rm{SM}}$}
	& $\mathcal{L}$=500 fb$^{-1}$ & 160 & 0.018 & 0.811 & 0 \\
	& $\mathcal{L}$=100 fb$^{-1}$ & 35  & 0.458 & 0.458 & 0.013 \\
\hline
\multirow{2}{*}{KK}
	& $\mathcal{L}$=500 fb$^{-1}$ & 971 	& 0.724	& 0.011 	& 0.18 \\
	& $\mathcal{L}$=100 fb$^{-1}$ & 181 	& 0.933 	& 0.822 	& 0.999 \\
\hline
\hline
\end{tabular}
\label{table:angularKolmogorov}
\end{table}
\begin{table}[!th]
\centering
\caption{\textsl{The unbinned Kolmogorov test for the $\sqrt{\hat s}$ distributions within $1 \leq \sqrt{\hat s} \leq 6$ TeV.}}
\vspace{4mm}
\begin{tabular}{ccc||ccc}
\hline\hline
	MC ref'  	& Luminosity		& \#Events of		& SM				& $Z'_{\rm{SM}}$ 	& KK\\
	model 			& of pseudo-data 	& pseudo-data	& pseudo-data	& pseudo-data	& pseudo-data\\
\hline\hline
\multirow{2}{*}{SM}
	& $\mathcal{L}$=500 fb$^{-1}$ & 2400 & 0.254 & 0.002 	& 0 \\
	& $\mathcal{L}$=100 fb$^{-1}$ & 480	 & 0.887 & 0.006 	& 0 \\
\hline
\multirow{2}{*}{$Z'_{\rm{SM}}$}
	& $\mathcal{L}$=500 fb$^{-1}$ & 2300 & 0.017		& 0.663 & 0 \\
	& $\mathcal{L}$=100 fb$^{-1}$ & 460	 & 0.28 		& 0.016 & 0 \\
\hline
\multirow{2}{*}{KK}
	& $\mathcal{L}$=500 fb$^{-1}$ & 1980 & 0 & 0 & 0.308 \\
	& $\mathcal{L}$=100 fb$^{-1}$ & 400	& 0  & 0 & 0.108 \\
\hline
\hline
\end{tabular}
\label{table:mHatKolmogorov}
\end{table}

In light of the results seen in Tables~\ref{table:angularKolmogorov} and~\ref{table:mHatKolmogorov}, it should be mentioned that a binned Kolmogorov test can provide better results ({\em ie}, the models will be more self-compatible and less inter-compatible), depending on the arbitrary choice of the binning.

\subsubsection{A summary of the overall procedure}
\begin{enumerate}
\item Looking at Table~\ref{table:angularFit} and comparing the values of $A_0$ and $A_{fb}$ around the expected resonance, it can be concluded that:
\begin{itemize}
\item Based on the fit results for the $A_0^{{\rm{CS}}}$ coefficients, for the three models (in the CS $\mathcal{O}'$ frame), they turn out to be consistent with zero.
\item At $\mathcal{L}$=100 fb$^{-1}$, the values of $A_{fb}^{{\rm{CS}}}$ for the three models are compatible with their MC reference estimations within less than one sigma.
\item At $\mathcal{L}$=500 fb$^{-1}$, the values of $A_{fb}^{{\rm{CS}}}$ for the SM and the $Z'_{\rm{SM}}$ models are compatible with their MC reference estimations within less than one sigma, where the KK model is compatible only within 2.3 sigma.
\item The errors for $A_{fb}^{{\rm{CS}}}$ in both luminosities are too high.
Hence, the sensitivity for probing the couplings using the measured $A_{fb}^{{\rm{CS}}}$ coefficients requires larger samples.
\end{itemize}

\item From the unbinned Kolmogorov tests in Table~\ref{table:angularKolmogorov} for the $\cos\theta '$ distributions, it can be concluded that:
\begin{itemize}
\item There is clear distinction between the KK model (pseudo-data) and the $Z'_{\rm{SM}}$ model (MC ref') at $\mathcal{L}$=500 fb$^{-1}$ as well as at $\mathcal{L}$=100 fb$^{-1}$.
\item The $Z'_{\rm{SM}}$ model (pseudo-data) and the KK model (MC ref') are compatible at $\mathcal{L}$=100 fb$^{-1}$.
However, the KK pseudo-data represents better its MC ref'.
\end{itemize}

\item From the unbinned Kolmogorov tests in Table~\ref{table:mHatKolmogorov} for the $\sqrt{\hat s}$ distributions, it can be concluded that:
\begin{itemize}
\item As expected, there is a clear compatibility between all three models (pseudo-data) to the data simulated with the equivalent at $\mathcal{L}$=500 fb$^{-1}$ and even at $\mathcal{L}$=100 fb$^{-1}$ except for the $Z'_{\rm{SM}}$ model at $\mathcal{L}$=100 fb$^{-1}$ which is very small ($\sim$0.016)
\item There is a significant distinction between the KK model and the $Z'_{\rm{SM}}$ model at both luminosity values.
\end{itemize}

\end{enumerate}
The important conclusion is that assuming an observed spin-1 resonance at $\sqrt{\hat s} \simeq $ 4 TeV, above the SM background, it will be possible to distinguish between the KK and $Z'_{{\rm{SM}}}$ models based on the Kolmogorov test for the $\sqrt{\hat s}$ distributions, already at $\mathcal{L}$=100 fb$^{-1}$.
For the $\cos\theta '$ distributions, the Kolmogorov test and the ML fit for the forward-backward asymmetry measurement can be very important in providing supportive information for such a discrimination.
This conclusion is valid already at $\mathcal{L}$=100 fb$^{-1}$ but it is much stronger for $\mathcal{L}$=500 fb$^{-1}$.
In addition, it is clear that for both luminosities, the measurement of $A_{fb}^{{\rm{CS}}}$ is too coarse for placing a precision statement about the new exotic couplings and it is only at higher integrated luminosities where a more sensitive study can be performed.


\section{Conclusions and Outlook}
In this paper the scenario of an observed resonance around 4~TeV, arising from the measurement of di-muon events is discussed.
It is shown that for the LHC, it will be possible to discriminate between the specific KK model and the $Z'_{{\rm{SM}}}$ model described in this paper assuming collisions in the design energy $\sqrt{s}=14$~TeV and assuming integrated luminosity of $\mathcal{L}$=100~fb$^{-1}$.
This statement relies both on the measurement of the $\sqrt{\hat s}$ and $\cos\theta '$ distributions and the measurement of the forward-backward asymmetry, applying as few cuts as necessary to deal with ISR.
One should keep in mind that this conclusion can not be complete without passing the generated events through the full $ATLAS$ detector simulation.
In that context, it should also be commented that:
\begin{itemize}
\item It is mandatory to apply a full detector simulation in order to treat the Kolmogorov unbinned (single) test as a single experiment.
Furthermore, one needs to repeat these experiments with large number of generated pseudo-data samples to quantify the sensitivity of the test.
\item The comparison should also be done with a spin-2 model (namely, the RS graviton).
In that context, the azimuthal angle distribution can be used as well\cite{AZIMUTHALRS}.
\item A measurement of both di-muon and di-electron events will double the statistics shown here.
\item Since the number of events is linear to the integrated luminosity, then it is clear that a sample much smaller than $\mathcal{L}$=100~fb$^{-1}$ will not be sufficient for any discrimination.
\item In the first year of the LHC operation, it is expected that the collisions will take place at $\sqrt{s}=$7-10~TeV CM energy and the luminosity is not expected to exceed 0.2-0.3~fb$^{-1}$.
In this case it will not be possible to discover in the first year a resonance at 4~TeV related to any of the models discussed in this paper.
\item For the possibility of no observed resonance below the current lower bound on $m^*$ ($\sim$4~TeV for KK), one should be able to place a new lower bound on its mass as demonstrated in~\cite{EVGENYTHESIS}.
\item Finally, the unique behavior of the invariant mass distribution of the KK model below the resonance may provide hints for the existence of such a resonance even if it is beyond the LHC reach.
\end{itemize}

The preliminary analysis presented here is based on results from the combination of Pythia8 and the new \mosesWithSpace software.
The specific Kaluza-Klein model is now implemented in \mosesNoSpace.
As declared, this is the first phase towards integration inside Pythia8 as an internal process.

Other plans include interfacing this generator to the $ATLAS$ detector simulation software for continuing the systematic analysis, generalize and re-order this framework as can be expected from the next releases and, enhance this framework with more BSM processes.

\section{Acknowledgments}
Our thanks are given to Y. Oz, and E. Yurkovsky (\textit{Group of exotic physics for $ATLAS$ analysis, The Raymond and Beverly Sackler School of Physics \& Astronomy, Tel Aviv University}).
Special thanks are given to Torbj\"{o}rn Sj\"{o}strand (\textit{Department of Theoretical Physics, Lund University}) for his support in this project.
Finally, for enabling the best conditions for the work done at UCL, the authors would like to thank MCnet and especially M. Seymour and J. Butterworth.

\clearpage


\appendix
\appendixpage
\addappheadtotoc

\section[Derivation of the KK tower for a $5d$ real massless scalar field]{Derivation of the KK tower for a $\boldsymbol{5d}$ real massless scalar field}
To demonstrate in practice the origin of the so called KK tower and the mass relation to the size of the parallel ED, $R$, we can start by thinking of a toy model where we have a massless scalar field in flat $5d$ space-time.
The metric tensor is $g_{AB} = (1,-1,-1,-1,\pm 1)$, assuming a space-like ED where we denote the ED coordinate by $z$ with $x$ being the regular four-coordinate~\cite{KKTOWERS,RIZZOPEDAGOGICAL}.

First, we assume that the field $\Phi(x,z)$ satisfies the $5d$ Klein-Gordon (KG) equation
\begin{equation}
\left( \partial_A \partial^A \right) \Phi(x,z) = \left( \partial_\mu \partial^\mu - \partial_z^2 \right)\Phi(x,z) = 0
\label{eq:5dKleinGordon}
\end{equation}
A second assumption, followed from the form of the KG equation, is to separate the dependencies (of the $x$ and $z$ coordinates) and take $\Phi(x,z)=\sum\limits_{n=0}^\infty{\phi_n(x) \psi_n(z)}$.
The next assumption is that $\psi_n(z)$ satisfies the equation
\begin{equation}
\partial_z^2 \psi_n(z) = -m_n^2 \psi_n(z)
\label{eq:5thCoordinateFunction}
\end{equation}
We also assume the orthogonality of the different $z$ dependent parts {\em ie},
\begin{equation}
\int\limits_{z_1}^{z_2}{dz \, \psi_m(z) \, \psi_n(z)} = \delta_{mn}
\label{eq:5dorthogonality}
\end{equation}
and finally, we write the $5d$ action of the field with the boundary conditions (BC) taken to be $\left[\psi_k(z) \partial_{z} \psi_n(z) \right]_{z_1}^{z_2} = 0$
\begin{equation}
\mathcal{S} = \int{d^4 x}\int\limits_{z_1}^{z_2}{dz \frac{1}{2} \left( \partial_A \Phi \partial^A\Phi \right)}
\label{eq:5daction}
\end{equation}\\
By integrating over the extra coordinate $z$, we reduce the $5d$ action to an effective $4d$ one
\begin{equation}
\mathcal{S} = \sum\limits_{n=0}^{\infty} \int{d^4 x \frac{1}{2}{\left[ \partial_\mu \phi_n \partial^\mu \phi_n - m_n^2 \phi_n^2 \right]}}
\label{eq:4daction}
\end{equation}
This action can be simply understood as the sum of distinct actions of $4d$ scalar fields $\phi_n(x)$ with different masses labeled by the integer index $n$.
In fact, the various masses we observe in $4d$ correspond to quantized values of the momentum along the extra coordinate $z$ since
\begin{equation}
\begin{array}{rl}
	&0 = p^2=g_{AB} p_A p^B = p_\mu p^\mu - p_z^2\\
	&p_\mu p^\mu = m_{4d}^2
\end{array}
\label{eq:momentum2}
\end{equation}
where $m_{4d}$ is the "observed" mass in $4d$.
The quantized momentum in the ED is $p_z = i\partial_z$ so by using Eq~\ref{eq:5thCoordinateFunction} we see that
\begin{equation}
m_{4d}^{(n)} = m_n
\label{eq:m4dm5dmn}
\end{equation}
where it is more appropriate to identify also the mass in $4d$ with the integer index $n=0,1,2,...$.
The last term is often called the KK tower of excitations. This is a striking result since we obtained a mass property in $4d$ by starting from a flat, $5d$ space-time with a massless field.
Yet, this doesn't explain the origin of the mass relation to the size of the extra dimension $R$. To obtain this relation, we have to look at the specific ED topology and the BC.
Looking on Eq~\ref{eq:5thCoordinateFunction}, we can expand $\psi_n(z)$ in exponential wave-eigenfunctions
\begin{equation}
\psi_n(z) = C_n^+ e^{i m_n z} + C_n^- e^{-i m_n z}
\label{eq:harmonic}
\end{equation}
where $C_n^\pm$ are constants.
If we postulate that the ED is compact such that it is curled into a one-dimensional sphere $S^1$ of radius $R$ then the specific translation invariance $z \to z + 2 \pi R$ of $\psi_n(z)$ implies that the BC are periodical {\em ie},
\begin{equation}
\psi(\pi R) - \psi(-\pi R)=0 \,\,\,\,\Leftrightarrow\,\,\,\, e^{2\pi i R m_n} - 1 = 0 \,\,\,\,\Leftrightarrow\,\,\,\, m_n = \frac{n}{R}
\label{eq:mass=radius}
\end{equation}
Therefore, Eq~\ref{eq:m4dm5dmn} translates to $m_{4d}^{(n)} = \frac{n}{R}$ where we see that the massless mode does exist as required, {\em ie}, $m_{4d}^{0}=0$.
However, in Eq~\ref{eq:kk_mass} we saw that the mass we observe in $4d$ consists of the KK excited term $\frac{n}{R}$ but also from another fixed term $m_0$ which is identified as the $4d$ state. We know that this term can be non-zero and that it is the Higgs mechanism which is responsible for creating the $4d$ mass $m_0$.
Thus, we obtained the full expression for the KK tower in this specific topology.
\begin{equation}
m_{4d}^{(n)} = \sqrt{m_0^2 + \left(\frac{n}{R}\right)^2}
\label{eq:fullmasstower}
\end{equation}

If we also define a parity operation on the interval $z \in \left[-\pi R, \pi R\right]$ then we obtain the $z \to -z$ mapping. This implies that there are two special, fixed points $z=0,$ and $z=\pi R$ which are left invariant by this $Z_2$ operation when combined with the periodicity property. The eigenfunctions that build $\psi_n(z)$ must now respect the discrete $Z_2$ parity symmetry so the solution for $\psi_n(z)$ given in Eq.(\ref{eq:harmonic}) has to be modified to either an even solution $A_n \cos\left(\frac{n}{R} z\right)$ or and odd one $B_n \sin\left(\frac{n}{R} z\right)$. The massless mode can belong now only to the even solution. This geometry is called the $S^1/Z_2$ orbifold.

The scalar field derivation is of course merely a single example of some illustrative methodological steps and assumptions concluding the origin of the KK tower and the mass relation to the ED radius $R$.

\clearpage

\section{The M{\small OSES} software}

\subsection{Download, configure and build}
The source code can be viewed at \href{http://projects.hepforge.org/moses}{http://projects.hepforge.org/moses},
or downloaded from \href{http://www.hepforge.org/downloads/moses}{http://www.hepforge.org/downloads/moses} for the latest version.
To prepare for installation, make sure to have LHAPDF, ROOT, HepPDT and Pythia8 installed (Pythia8 should be configured with {\tt --enabled-shared}).
Make sure that the scripts root-config and lhapdf-config are in your {\tt PATH} (located in the bin directories of these packages).
Copy the full paths to HepPDT and Pythia8 base directories, these paths will need to be specified during configuration.
To configure and build \mosesNoSpace,
\begin{verbatim}
	wget http://www.hepforge.org/archive/moses/moses-<x>.<y>.<z>.tgz
	tar -zxvf moses-<x>.<y>.<z>.tgz
	cd moses-<x>.<y>.<z>
	source configure.sh --heppdtpath=/path/to/heppdt
                    --pythia8path=/path/to/pythia8
	make
\end{verbatim}
To clean the installation of compiled objects, run {\tt make clean}.
To complete purge of transient files, and to unset all the exported variables, run {\tt make cleanall}.
To unset all the exported variables, {\tt source configure.sh --unsetal}.
On the next logons, after the initial configuration, one has to run {\tt source ./config.sh} to set the necessary variables for runtime and for recompilations.

\subsection{Basic usage}
All the executables go in the {\tt bin/execs} directory and they can be ran directly from there.
To see the list of executables, run {\tt ls -lrt bin/execs}.
To run a certain executable ({\tt executable\_name}), run {\tt \$MOSESSYS/bin/executable\_name}.

There is a special {\tt helper} script for handling with running and re-making the various executables.
To see all the {\tt helper} script options, run {\tt ./helper --help}.
To see the list of executables, run {\tt ./helper --ls}.
To run some basic examples, one can execute
{\tt ./helper --run=usageExample\_BuildProcess} or {\tt --run=usageExample\_LHAPDF} or {\tt --run=usageExample\_HepPDT}.

All the output is essentially in the form of, either flat files ({\tt .dat} etc.), or {\tt .root} files (trees or canvases) or {\tt .eps} files.
All the executables output go by default to, {\tt data/}, where this is determined automatically by the variable,
{\tt \$MOSESDATA/ = \$MOSESSYS/data/ = /full/path/to/data/}.
This can be overridden by reexporting this variable,
\begin{verbatim}
mkdir /your/path/to/the/new/datadir
export MOSESDATA=/your/path/to/the/new/datadir
\end{verbatim}
Possibly, include the last line in the end of the {\tt config.sh} script which has to be executed once on every logon after the initial installation.
To see the data in its default location, run {\tt ls -lrt data/} or in general, {\tt ls -lrt \$MOSESDATA/}, where the later should be used in the case of a user defined location.

\subsection{Structure}
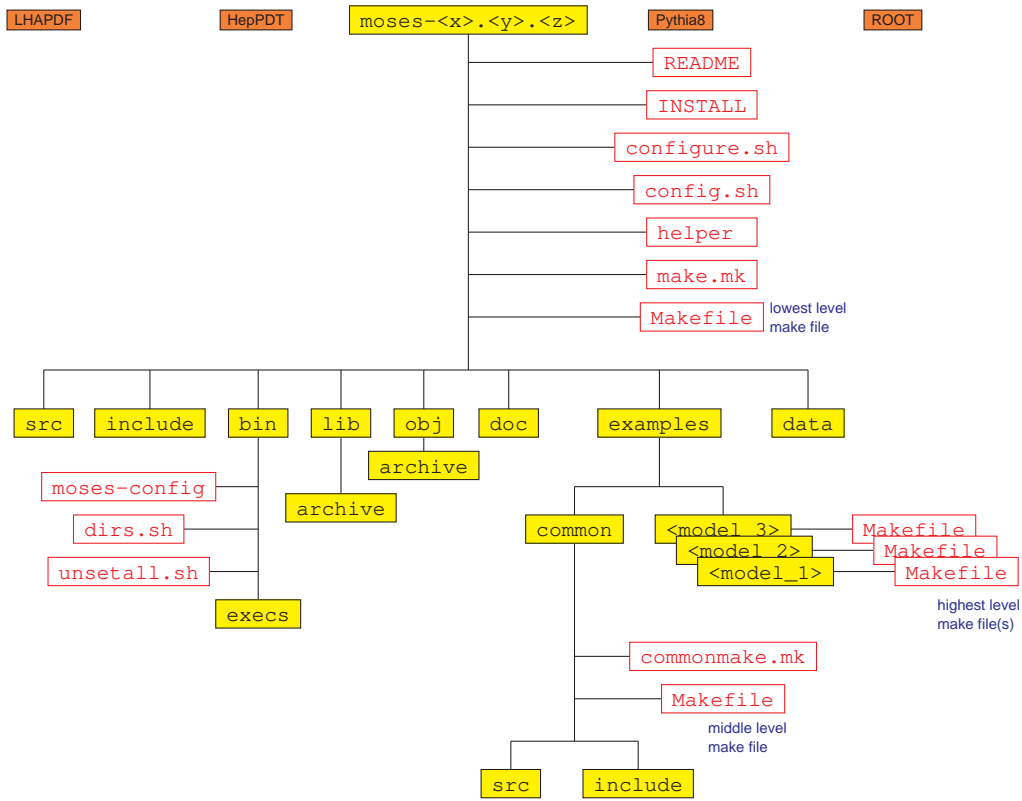
\begin{figure}[!th]
\begin{center}
\begin{picture}(320,320)(0,0)
\SetScale{0.8}

\SetColor{Black}
\Line(200,360)(200,195)
\Line(200,340)(310,340)
\Line(200,320)(310,320)
\Line(200,300)(310,300)
\Line(200,280)(310,280)
\Line(200,260)(310,260)
\Line(200,240)(310,240)
\Line(200,220)(310,220)

\Line(0,195)(360,195)
\Line(0,195)(0,170)
\Line(50,195)(50,170)
\Line(101,195)(101,170)
\Line(140,195)(140,170)
\Line(179,195)(179,170)
\Line(219,195)(219,170)
\Line(290,195)(290,170)
\Line(360,195)(360,170)
\Line(179,170)(179,150)
\Line(140,170)(140,130)
\Line(101,170)(101,80)
\Line(101,140)(40,140)
\Line(101,120)(40,120)
\Line(101,100)(40,100)
\Line(290,170)(290,140)
\Line(250,140)(320,140)
\Line(250,140)(250,120)
\Line(320,140)(320,120)
\Line(320,120)(410,120)
\Line(330,110)(420,110)
\Line(340,100)(430,100)
\Line(250,120)(250,20)
\Line(220,20)(280,20)
\Line(220,20)(220,0)
\Line(280,20)(280,0)
\Line(250,60)(320,60)
\Line(250,40)(320,40)

\SetPFont{Courier}{10}
\SetColor{Black}
\IfColor{\CText(200,360){Black}{Yellow}{moses-<x>.<y>.<z>}}{\GText(200,360){0.9}{moses-<x>.<y>.<z>}}
\IfColor{\CText(0,170){Black}{Yellow}{src}}{\GText(0,170){0.9}{src}}
\IfColor{\CText(50,170){Black}{Yellow}{include}}{\GText(50,170){0.9}{include}}
\IfColor{\CText(101,170){Black}{Yellow}{bin}}{\GText(101,170){0.9}{bin}}
\IfColor{\CText(140,170){Black}{Yellow}{lib}}{\GText(140,170){0.9}{lib}}
\IfColor{\CText(179,170){Black}{Yellow}{obj}}{\GText(179,170){0.9}{obj}}
\IfColor{\CText(219,170){Black}{Yellow}{doc}}{\GText(219,170){0.9}{doc}}
\IfColor{\CText(290,170){Black}{Yellow}{examples}}{\GText(290,170){0.9}{examples}}
\IfColor{\CText(360,170){Black}{Yellow}{data}}{\GText(360,170){0.9}{data}}
\IfColor{\CText(179,150){Black}{Yellow}{archive}}{\GText(179,150){0.9}{archive}}
\IfColor{\CText(140,130){Black}{Yellow}{archive}}{\GText(140,130){0.9}{archive}}
\IfColor{\CText(101,80){Black}{Yellow}{execs}}{\GText(101,80){0.9}{execs}}
\IfColor{\CText(250,120){Black}{Yellow}{common}}{\GText(250,120){0.9}{common}}
\IfColor{\CText(320,120){Black}{Yellow}{<model_3>}}{\GText(320,120){0.9}{<model_3>}}
\IfColor{\CText(330,110){Black}{Yellow}{<model_2>}}{\GText(330,110){0.9}{<model_2>}}
\IfColor{\CText(340,100){Black}{Yellow}{<model_1>}}{\GText(340,100){0.9}{<model_1>}}
\IfColor{\CText(220,0){Black}{Yellow}{src}}{\GText(220,40){0.9}{src}}
\IfColor{\CText(280,0){Black}{Yellow}{include}}{\GText(280,40){0.9}{include}}

\SetPFont{Courier}{10}
\SetColor{Red}
\BText(310,340){README}
\BText(310,320){INSTALL}
\BText(310,300){configure.sh}
\BText(310,280){config.sh}
\BText(310,260){helper }
\BText(310,240){make.mk}
\BText(310,220){Makefile}
\BText(40,140){moses-config}
\BText(40,120){dirs.sh}
\BText(40,100){unsetall.sh}
\BText(320,60){commonmake.mk}
\BText(320,40){Makefile}
\BText(410,120){Makefile}
\BText(420,110){Makefile}
\BText(430,100){Makefile}

\SetPFont{Helvetica}{7}
\SetColor{Blue}
\PText(342,222)(0)[lb]{lowest level}
\PText(342,213)(0)[lb]{make file}
\PText(313,24)(0)[lb]{middle level}
\PText(313,15)(0)[lb]{make file}
\PText(421,82)(0)[lb]{highest level}
\PText(421,73)(0)[lb]{make file(s)}

\IfColor{\CText(0,360){Black}{Orange}{LHAPDF}}{\GText(0,360){0.9}{LHAPDF}}
\IfColor{\CText(100,360){Black}{Orange}{HepPDT}}{\GText(100,360){0.9}{HepPDT}}
\IfColor{\CText(300,360){Black}{Orange}{Pythia8}}{\GText(300,360){0.9}{Pythia8}}
\IfColor{\CText(400,360){Black}{Orange}{ROOT}}{\GText(400,360){0.9}{ROOT}}

\end{picture}
\caption{\textsl{The \mosesWithSpace software structure after installation. The basic compilation flow is from the lowest level Makefile (in {\tt moses-<x>.<y>.<z>}) to the middle level Makefile (in {\tt moses-<x>.<y>.<z>/examples/common}) and to the highest level Makefile(s) (in {\tt moses-<x>.<y>.<z>/examples/model\_i}).}}
\label{fig:MosesStructure}
\end{center}
\end{figure}

In general, the basic structure is very simple so it can be very easy to either run the examples and to allow further personal development on top of the existing structure.
The common area (under the {\tt moses-<x>.<y>.<z>/examples}) contains common code so it can be linked against any package (also under the {\tt examples} directory).
Each package (one directory) should correspond to a single model.
In other words, a single hard process with the specific related tasks to be executed (except for the tasks that can be thought of as common to several packages).
The user can copy code from these two areas for his own needs, add new libraries or add new packages just like it is done in the examples directory.
For a detailed picture of the \mosesWithSpace software structure, see Fig~\ref{fig:MosesStructure}.
The compilation flow is organized from the lowest level (core) to the highest level (models), so when standing in one location which contains a \shellfont{Makefile} and executing ``make'', everything that is at higher levels will also compile if changed.
The {\tt moses-<x>.<y>.<z>/make.mk} and {\tt moses-<x>.<y>.<z>/examples/common/commonmake.mk} files contain global definitions at different levels to avoid rewriting these definition in each of the other Makefiles.
Thus, either one or two of theses files have to be included in every Makefile depending on the context.
Note that there can be many tasks placed under each model directory.
That includes oriented plugins and/or main programs like MC-programs (pythia8-mains, pythia8-plugins), theoretical-analysis programs, data-analysis programs etc.
Therefore, there are some places where one can find specific higher levels than the ones sketched in Fig~\ref{fig:MosesStructure}.
In particular there are analysis directories that encapsulate several analysis tasks, some of which presented in this paper.

\end{document}